\documentclass{vgtc}                          % final (conference style)
\graphicspath{{figures/}{pictures/}{images/}{./}} % where to search for the images

\usepackage{times}                     % we use Times as the main font
\usepackage{tabu}                      % only used for the table example
\usepackage{booktabs}                  % only used for the table example
\usepackage{lipsum}                    % used to generate placeholder text
\usepackage{mwe}                       % used to generate placeholder figures
\usepackage{amsthm}

\usepackage{paralist}
\usepackage{mathptmx}                  % use matching math font
\usepackage{amsmath}
\usepackage{amssymb}

\onlineid{5966}

\vgtccategory{Research}

\vgtcinsertpkg

\usepackage{enumitem}
\usepackage{complexity}

\usepackage[ruled, linesnumbered]{algorithm2e}
\usepackage{subcaption}
\usepackage{complexity}

\newcommand{\mypar}[1]{\smallskip\noindent\textbf{#1}}
\newcommand{\TLL}[0]{\ell}
\newcommand{\sites}[0]{\mathcal{F}}
\newcommand{\site}[0]{\mathcal{s}}
\newcommand{\ports}[0]{\mathcal{P}}

\usepackage{complexity}
\newcommand{\attributionMapData}{background map obtained from \url{maputnik.github.io} with \copyright 2015 Lukas Martinelli \copyright 2024 MapLibre contributors under MIT license}

\newcommand{\attributionMapDatatwo}{background map obtained from \url{commons.wikimedia.org} with \copyright 2010 Uwe Dedering under  CC BY-SA 3.0 license}

\newcommand{\revOld}[1]{\textcolor{black}{#1}}
\newcommand{\rev}[1]{\textcolor{black}{#1}}
\newcommand{\del}[1]{}

\newcommand{\osf}[0]{\href{https://osf.io/3ekjb}{OSF} }

\Crefname{algocf}{Algorithm}{Algorithms}

\title{Clarity and Computational Efficiency of Orbital Boundary Labeling}

\author{
Markus Wallinger\thanks{e-mail: markus.wallinger@tum.de}\\ %
        \scriptsize Technical University of Munich %
\and Annika Bonerath\thanks{e-mail: bonerath@igg.uni-bonn.de}\\ %
     \scriptsize University of Bonn %
\and Soeren Terziadis\thanks{e-mail: sterziadis@ac.tuwien.ac.at}\\ %
     \scriptsize TU Eindhoven
\and Jules Wulms\thanks{e-mail: j.j.h.m.wulms@tue.nl}\\ %
     \scriptsize TU Eindhoven
\and Martin N\"ollenburg\thanks{e-mail: noellenburg@ac.tuwien.ac.at}\\ %
     \scriptsize TU Wien
}

\teaser{
  \centering
      \begin{subfigure}[b]{0.40\linewidth}
      	\centering
      	\includegraphics[width=.9\linewidth]{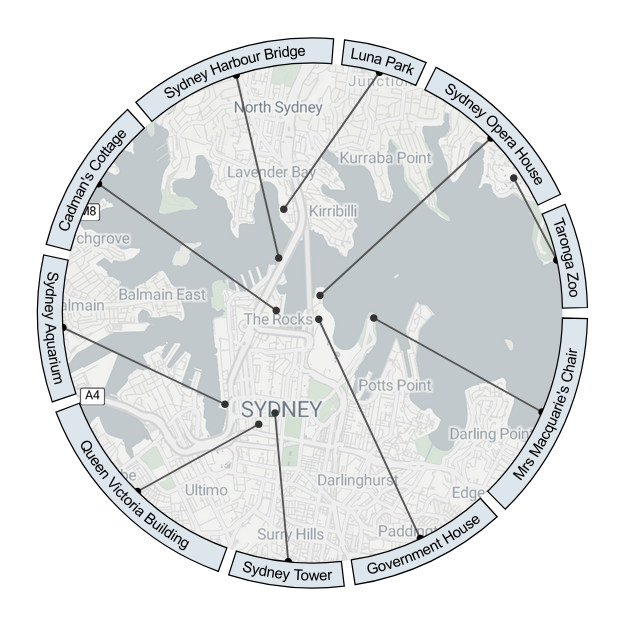}
      	\caption{Straight-line Leaders}
      	\label{fig:teaser_sl_nu}
      \end{subfigure}
      \qquad
      \begin{subfigure}[b]{0.40\linewidth}
      	\centering
      	\includegraphics[width=.9\linewidth]{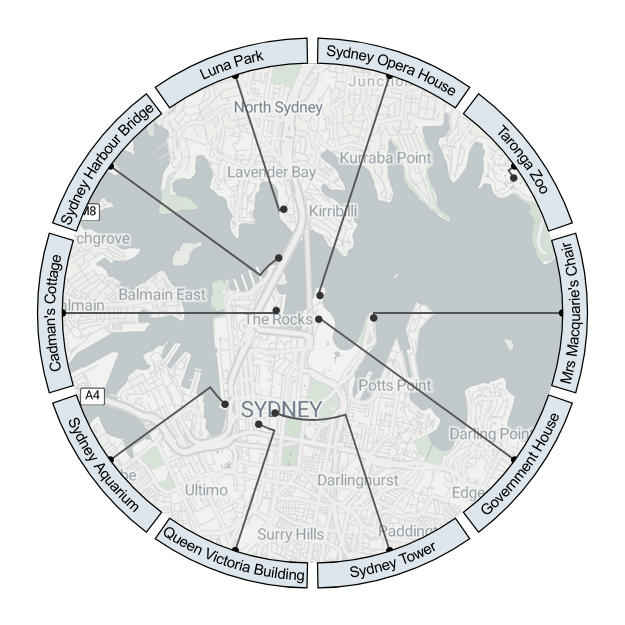}
      	\caption{Orbital-radial Leaders}
      	\label{fig:teaser_or_uni}
      \end{subfigure}%
  \caption{Illustrative example of our orbital labeling approach to annotate tourist attractions. \textbf{(a)} shows straight-line leaders for non-uniform label sizes. In \textbf{(b)} the same example is shown with uniform label sizes and orbital-radial leaders.}

  \label{fig:teaser}
}

\abstract{
\revOld{Circular interfaces such as those found on smartwatches, automotive dashboards, cockpit instruments, or in radial visualizations pose unique challenges for placing readable labels.
Traditional rectangular labeling methods waste screen space and create visual clutter on these constrained displays.}
In orbital boundary labeling, the labels (e.g., the features' names) are placed in an annulus-shaped orbit outside of the figure, 
and each label is connected to its feature using a short, crossing-free leader line.
We contribute algorithms to compute two leader styles, orbital-radial and straight-line, for uniform and non-uniform label sizes, optimizing for crossing-free shortest leaders. 
We evaluate the model and the algorithms with computational experiments and a controlled user experiment. The user experiment reveals that both leader types exhibit similar accuracy, but straight-line leaders yield faster response times.
All supplemental materials are available at \osf.
} % end of abstract


\begin{document}


\firstsection{Introduction}

\maketitle

The challenge of effectively annotating visual elements, e.g., in maps, diagrams, medical images, and technical illustrations, with textual or symbolic labels is central to many fields, including cartography, data visualization, and geographic information systems. \emph{External labeling}~\cite{BekosNN19}, where labels are placed along the boundary of a visualization and connected to features via leader lines, \revOld{as seen in \Cref{fig:teaser}}, is a well-studied approach to avoid visual clutter within the primary display area. 
Traditional methods typically employ rectangular labels along rectangular or circular boundaries. However, these conventional techniques may be insufficient for certain specialized layouts or constrained display geometries, particularly for circular displays and radial visualization designs.
%, which we investigate in this paper.

\begin{figure*}[t]
    \centering
    \begin{subfigure}{.2\linewidth}
        \centering
        \includegraphics[width=\linewidth,page=3]{figures/vis_designspace.pdf}
        \subcaption{Internal labeling}
        \label{fig:designspace_smartwatches1}
    \end{subfigure}
    \quad
    \begin{subfigure}{.2\linewidth}
        \centering
        \includegraphics[width=\linewidth,page=2]{figures/vis_designspace.pdf}
        \subcaption{Classic external labeling}
        \label{fig:designspace_smartwatches2}
    \end{subfigure}
    \quad
    \begin{subfigure}{.2\linewidth}
        \centering
        \includegraphics[width=\linewidth,page=1]{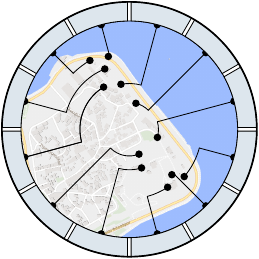}
        \subcaption{Orbital boundary labeling}
        \label{fig:designspace_smartwatches3}
    \end{subfigure}
    \quad
    \begin{subfigure}{.21\linewidth}
        \centering
        \includegraphics[width=\linewidth,page=5]{figures/mlb_v2.pdf}
        \subcaption{Thematic maps}
        \label{fig:designspace_cartogram}
    \end{subfigure}
    \caption{ (a)-(c) illustrates challenges on circular displays for labeling a map (\attributionMapData) while (d) shows a sketch of orbital boundary labeling for thematic maps (\attributionMapDatatwo).}
    \label{fig:designspace_smartwatches}
\end{figure*}

Small circular displays, such as those in smartwatches, present significant challenges for displaying spatial information on digital maps. Larger devices, such as tablets or laptops, can comfortably label numerous points of interest, providing users with a clear and uncluttered overview. In contrast, the smaller screens of smartwatches allow far fewer features to be labeled without obstructing the map. Although interactivity through panning and zooming offers a workaround, these actions are often inefficient and disrupt the user’s context due to the limited viewport size. The circular nature of many smartwatch displays adds another layer of complexity. While rectangular displays benefit from existing visualizations designed for standard viewports, circular screens face unique limitations.
Adapting rectangular visualizations to circular displays often results in wasted space and less cohesive designs, see \Cref{fig:designspace_smartwatches1,fig:designspace_smartwatches2}.
% See \Cref{fig:designspace_smartwatches1,fig:designspace_smartwatches2,fig:designspace_smartwatches3} for an illustration.
Despite their growing popularity, circular displays are rarely considered in visualization research, highlighting the need for approaches specifically tailored to their geometry. Existing techniques have addressed some of these limitations for rectangular displays~\cite{GedickeBNH21, DBLP:journals/jlbs/PerebnerHG19}, and we consider small circular displays as one application scenario for our technique.  
\del{Furthermore,} Larger circular displays are also used in cockpits and automotive dashboards. 
%Furthermore, larger circular displays, such as those used in cockpits and automotive dashboards, could also benefit from orbital boundary labeling, much like their smaller counterparts.

As a second example, radial layouts, which are frequently used in thematic maps, donut charts, and circular bar charts, present an alternative to rectangular designs. Such layouts are not only aesthetically pleasing~\cite{BurchW14} but also improve cognitive recall and memorability~\cite{BorkinVBISOP13}. Encoding information in thematic maps can be difficult, as areas can be rather small. Thus, designs such as ring maps~\cite{bslrm-rmsvmed-11} or necklace maps~\cite{sv-anm-15} have been proposed to visualize information outside of the primary map view. This paper presents an alternative by connecting each feature to a glyph using a leader. An illustration is shown in \Cref{fig:designspace_cartogram}.

Recently, Bonerath et al.~\cite{BonerathNTWW24} introduced orbital boundary labeling and investigated the theoretical aspects of several labeling variants. In this paper, we build upon this work and \revOld{extend orbital boundary labeling with the following contributions:}

\textbf{(1)} \revOld{We focus on those variants of orbital boundary labeling that are most relevant for visualization practice. In contrast to~\cite{BonerathNTWW24}, we exclude variants that are mainly of theoretical interest and introduce an additional simplification by fixing the alignment of the first label, which enables efficient heuristic computation.}
\textbf{(2)} \revOld{Prior work has shown that minimizing total leader length for non-uniform labels under crossing-free constraints is weakly \NP-hard~\cite{BonerathNTWW24}. We complement this result by introducing both exact and heuristic algorithms that optimize leader length in practice and by evaluating the trade-off between solution quality and runtime through computational experiments.} 
\textbf{(3)} In a user experiment, we compare different variants of orbital boundary labeling, measuring the task completion time and accuracy with which users can perform basic tasks.

\section{Related Work}
\label{sec:relatedwork}

%\textbf{MN: collected comprehensive list of relevant references on external labeling; small screen is currently just two papers, but from the survey, one may follow up on some more references}

Text labels and symbols have long been used to annotate features in illustrations, technical drawings, and especially maps. While much work in (automated) cartography has focused on optimizing the placement of \emph{internal} labels for readability~\cite{i-pnm-75,y-lal-72,Agarwal.1998,vanKreveld.1999,QuZCCL25} (see~\Cref{fig:designspace_smartwatches1}), i.e., labels placed directly next to visual features, this paper considers \emph{external} labeling~\cite{BekosNN19,bnn-elfcat-21}. In external labeling, labels are placed outside the illustration, connected to features by thin leader lines, which minimizes interference with the underlying image. Computationally, internal labeling is commonly related to NP-hard geometric independent set problems~\cite{ChristensenMS95,FormannW91}, while external labeling focuses primarily on optimizing crossing-free, minimal-length leaders for labels arranged around the illustration.

In \emph{external labeling}, approaches can be characterized by the geometry of the labeling boundary (e.g., rectangular, circular), leader shape (straight, orthogonal, Bézier, etc.), and label properties (uniform height, variable width). Some methods rely on heuristics to produce short, crossing-free leaders, while others compute provably optimal labelings under objectives such as minimizing total leader length (also known as ``ink''). For a recent survey and taxonomy of external labeling techniques, see Bekos et al.~\cite{BekosNN19,bnn-elfcat-21}.

The most established style is \emph{boundary labeling}, where labels are aligned to the sides of a rectangular bounding box. Bekos et al.~\cite{BekosKSW07} provided polynomial-time algorithms for various leader types (e.g., orthogonal, straight). This has since been extended to more general objectives~\cite{bhkn-amcbl-09} or leader geometries~\cite{bkns-blol-10}. Other work introduces sliding label positions along the edge~\cite{nps-dosbl-10,hpl-blflp-14} or suggests fast heuristics such as force-based methods~\cite{AliHS05} and simulated annealing~\cite{gah-aepspfsmer-23}. Extending beyond rectangular borders, Niedermann et al.~\cite{nnr-rclwsl-17} studied convex contour boundary labeling with angle-monotone straight leaders. Barth et al.~\cite{BarthGNN19} conducted a controlled user study on different leader types for uniformly sized labels and found that orthogonal leaders perform well overall, while diagonal leaders are aesthetically preferred. However, circular boundary settings (as addressed in this paper) were not part of their study.

Another relevant direction is \emph{excentric labeling}, where labels lie outside a circular focus region within a rectangular map, connected via straight or one-bend leaders~\cite{Fekete1999,Bertini2009,Heinsohn_2014,Niedermann2019}. In some variants, labels are placed along a circular boundary~\cite{fhssw-alfr-12,Haunert2014}, though they are still axis-aligned rectangles and not tailored for circular displays.

\emph{Ring maps}~\cite{bslrm-rmsvmed-11} are conceptually related to our setting: labels are radially aligned along a circular boundary with straight leaders. However, the authors only consider uniformly sized labels and do not provide an algorithm to optimize the ordering or prevent crossings. Necklace maps~\cite{SpeckmannV10,sv-anm-15} have been suggested as an underlying technique but do not guarantee optimality for ring visualization.

Among works focused on small screens, Zoomless Maps~\cite{GedickeBNH21} address the display of many labeled features on smartwatches, using interactive exploration. While their context involves rectangular screens and different interactions, several of their ideas could complement orbital labeling, as discussed in Section~\ref{sec:discussion}.

Finally, the closest related work is by Bonerath et al.~\cite{BonerathNTWW24}, who introduced the \emph{orbital boundary labeling} model. They explore several degrees of freedom, such as fixed or free label ordering, port positions, uniform and non-uniform labels, and different leader styles. Their results include polynomial-time algorithms or \NP-hardness proofs, depending on the chosen settings. We build upon this foundation, focus on variants most relevant to visualization and propose the exact and heuristic algorithms for \NP-hard cases.

\section{Orbital Labeling Variants in a Visualization Context}\label{sec:design_space}

\newcommand{\anglec}{\theta}
\newcommand{\distance}{r}

We briefly introduce the orbital boundary labeling problem as stated in the work by Bonerath et al.~\cite{BonerathNTWW24} and discuss which variants have the most relevance in a visualization context.

\begin{figure}
	\centering
	\includegraphics[page=5,width=0.5\linewidth]{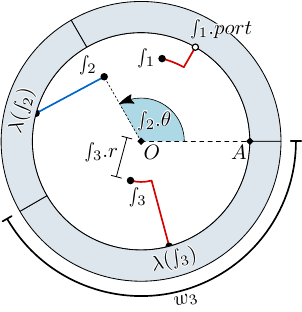}
	\caption{Illustration of the notation used in this paper. Note that both OR-leaders (red) and SL-leaders (blue) are illustrated, while any labeling we consider will exclusively contain one or the other.}
	\label{fig:notation}
\end{figure}

\mypar{Problem Description.}
\label{sec:probDesc}
The notation is illustrated in Figure~\ref{fig:notation}.
We are given a set $\mathcal{F} = \{\site_1, \dots, \site_n\}$ of point features.
The features are located within a disk with center~$O$, boundary~$B$ and radius~$R$.
Whenever we specify coordinates, we consider~$O$ to be the center of the coordinate system and we call the intersection point of a horizontal line starting at $O$ with $B$ the \emph{anchor}~$A$.
We assume that there is always one label whose circular arc along $B$ starts at $A$.
\revOld{Note that any other choice of $A$ would work, e.g., the bottom or top.}
A feature $\site_i$ is given in polar coordinates, i.e., the distance $\site_i.\distance$ from $\site_i$ to $O$ and the counter-clockwise angle $\site_i.\anglec = \measuredangle AO\site_i$.
%(see also Figure~\ref{fig:notation}). 
We will write $\vert B \vert$ for the length of~$B$.
A point feature has an associated label $\lambda(\site_i)$.
A label $\lambda(\site_i)$ is a circular arc along~$B$ of length~$w_i$, s.t., $\vert B\vert = \sum_{i=1}^n w_i$.
The output is a set of $n$ ports $\site_1.port, \ldots, \site_n.port$, s.t., the $\lambda(\site_i)$ can be placed with $\site_i.port$ at its center, all labels are pairwise interior-disjoint, and every $\site_i$ can be connected to the port at the center of $\lambda(\site_i)$ with a leader curve of a certain style.
In any valid solution, all leader curves have to be pairwise non-intersecting, and a solution is optimal if the total length of all leader curves is minimal over all valid solutions.
We refer to this as a \emph{leader length-optimal labeling}.
\revOld{Such an optimal solution has the advantage of requiring less ink, thus reducing the clutter.}

% properties of a labeling like 
Note that small gaps can be introduced between labels when rendering the labeling.
Finally, we assume a variant of general position for the placement of feature points, specifically for any pair of points $\site_i, \site_j$ we assume $\site_i.\distance \neq \site_j.\distance$. 
This assumption is made to avoid ambiguity between leaders and can be achieved by small perturbations to the feature points.
% This is a sensible assumption, as otherwise it is difficult to unambiguously associate leaders with two features at the same position.

\mypar{Variants.}
The four variants we consider are characterized by their leader style (SL vs.\ OR) and their label sizes (uniform vs.\ non-uniform).
%of the above problem definition, depending on two (binary) variables.
%First, we investigate two different leader styles: straight-line SL- and orbital-radial OR-leaders.
A \emph{straight-line} (SL) leader for a point feature~$\site_i$ is a line segment with endpoints~$\site_i$ and~$\site_i.port$, and its length is the Euclidean distance between these two endpoints.
In contrast, an \emph{orbital-radial} (OR) leader consists of an orbital circular arc concentric with $B$ starting at $\site_i$ and ending at the start point of a radial line segment (its supporting line contains $O$), which ends at~$\site_i.port$.
%Note that for a fixed position of~$\site_i$ and~$\site_i.port$ there are exactly two possible OR-leader.
The length of an OR-leader is the sum of the length of the circular arc segment and that of the radial line segment.
% Second, we distinguish variants based on the sizes~$w_i$ of the labels.
While SL-leaders for a chosen port are unique, OR leaders could run clockwise or counterclockwise; 
to avoid long leaders
% since we are minimizing leader length 
we always use the shorter of the two possibilities.
Label sizes are distinguished as follows.
If all labels have the same size $w_i = \vert B\vert / n$, then we say the labels are \emph{uniform}, otherwise they are \emph{non-uniform}.
%This yields four problem variants.

\mypar{Assumptions.}
% In this paper, we focus on variants that incorporate 
By varying label size and leader style, we consider two of the five problem space dimensions discussed in~\cite{BonerathNTWW24}. While the algorithmic feasibility is explored for all dimensions in~\cite{BonerathNTWW24}, here we focus on the variants that are expected to have the largest impact on usability in a visualization context.
% consider vary only in the considerations that we believe impact usability most orbital boundary labeling in a practical visualization context.
The three dimensions that we do not vary are the following.
\textbf{(1)} We always allow reordering of the labels, since a fixed label order can limit flexibility, potentially making it impossible to compute a crossing-free labeling.
\textbf{(2)} Ports are always placed in the center of a label to provide more predictable and uniform leader line attachments. While non-fixed port positions could slightly reduce overall leader length, the improvement is marginal and may not justify the added complexity.
\textbf{(3)} While we do not restrict the position of each label to one of a fixed set of candidate positions around $B$,
% we consider a slightly modified variant regarding fixed or non-fixed \textit{label positions}. Specifically, 
the anchor $A$ enforces a fixed starting point. For uniform labels, this fixes all label positions, but it has a limited impact on variants with non-uniform label sizes. This modification simplifies computation and is expected to have only a minimal effect on the total leader length.
% First, assuming a predetermined \textit{fixed order} of labels has some merit, as it enables sorting labels lexicographically or by value. However, a fixed label order can limit flexibility, potentially making it impossible to compute a crossing-free labeling.
% Second, using \textit{fixed port positions} provides more predictable and uniform leader line attachments. While non-fixed port positions could slightly reduce overall leader length, the improvement is marginal and may not justify the added complexity.
% Finally, we consider a slightly modified variant regarding fixed or non-fixed \textit{label positions}. Specifically, the anchor $A$ enforces fixed positions for uniform labels but has a limited impact on variants with non-uniform label sizes. This modification simplifies computation and is expected to have only a minimal effect on the total leader length.
\revOld{Additionally, this avoids more complex or computationally expensive heuristics as we do not need to compute potentially multiple solutions. However, for non-uniform labels, one implication is that an unfortunate choice of anchor can lead to no existing planar solution. Although we didn't encounter this in our computational experiment, we provide an example in the supplemental material for illustration purposes.}

\mypar{Practical Considerations}
\revOld{Although orbital boundary labeling was initially motivated by round smartwatch displays, the geometric assumptions of our model (a circular inner region with an annular label orbit) apply to a wide range of circular interfaces. Typical smartwatch screens (e.g., 440-480\,px in diameter) provide an outer circumference of roughly 1.4-1.5\,kpx, which comfortably accommodates 10--16 labels with short names in a readable font size (8--10\,pt). Hence, the annular orbit can be exploited effectively without compromising legibility. 
Beyond smartwatches, the same design is relevant for circular dashboards, cockpit dials, compact circular widgets, and radar- or sonar-like visualizations, where external labeling around a circular core reduces clutter and preserves the central display area. For such designs, the number of labels that we can realistically place is likely to be higher.}

\section{Labeling Algorithms}
\label{sec:algorithms}

In this section, we describe our algorithms for computing labelings for a given instance. We first explain the exact algorithms for uniform label size variants to compute leader length optimal solutions. We then introduce exact mathematical programming formulations as well as heuristics for variants with non-uniform label sizes.

% First, we introduce preliminaries for the problem on an abstract level. We assume that we are given a collection of sites $\sites = {\site_1, \dots, \site_n}$ where $n$ is the total number of sites. Let $O$, $B$ and $R$ be the center/origin of the coordinate system, the boundary of the circle and the radius of the circle respectively.\textcolor{red}{$R$ could simply be 1}
% We denote with $|B|$ the length of $B$.
% The Cartesian coordinates of a site are expressed as $\site_i.x$ and $\site_i.y$.
% Note that these coordinates can equivalently be expressed as polar coordinates consisting of the clockwise angle $\site.\angl$ between the point and the ray starting at $O$ and extending horizontally to the right  and a distance $\site.r$ from $O$.
% We frequently use the concept of polar coordinates in this section.
% Additionally, we assume that each site $\site_i$ has a weight $w_i$ assigned.
% Let $W = \sum_{i=1}^n w_i$.
% The label of site $\site_i$ is a radial arc of length $|B|\cdot w_i/W$ along $B$.
% We refer to its two endpoints as the clockwise and counter-clockwise endpoint (relative to $O$).
% We will refer to the port of the label of a site $\site_i$ as $\site_i.port$.
% Note that a port can also be expressed as a point in polar coordinates.
% All algorithms assume that anchor point is given, i.e., in any valid solution there is one label (the first label) whose clockwise endpoint has polar coordinates $(0, R)$.
% Note that the existence of this anchor point implies that a total order on the labels completely characterizes a solution.

\subsection{Uniform Labels}

\revOld{For completeness and as the algorithm is an essential building block of the heuristics, we give a more accessible overview of the algorithm here (see \Cref{alg:uniform}) than presented in Section 4.2 of Bonerath et al.~\cite{BonerathNTWW24}.} Our first observation is that the ports have fixed positions along boundary~$B$ relative to anchor~$A$: Since every label has the same size, the ports are exactly distance $w_i = |B|/n$ separated from each other.
%, and the two ports incident with~$A$ are at a distance~$w_i/2 = |B|/2n$ from $A$. 
As a first step in our algorithm, we compute these candidate positions for the ports.

\begin{algorithm}[t]
\caption{Orbital labeling with uniform labels}\label{alg:uniform}
\SetKwFunction{algo}{uniformLabeling}
\SetKwFunction{slcost}{computeCostSL}
\SetKwFunction{orcost}{computeCostOR}
\SetKwFunction{cost}{computeCost*}
\SetKwFunction{portfun}{computePorts}
\SetKwFunction{matching}{minimalWeightMatching}
\KwData{Point features $\sites$, $|\sites| = n$}
\KwResult{Port assignments for $\sites$}

   \SetKwProg{myalg}{Algorithm}{}{}
  \myalg{\algo{$\sites$}}{
    $\ports$ $\gets$ \portfun{n}\;
    $W$ $\gets$ \cost{$\sites, \ports$}; \, \tcp{* either SL or OR}
    $M$ $\gets$ \matching{$W$}\;
 
    \For{$\site \in \sites$}{
        $\site.port$ $\gets$ $M[\site]$\;
    }
    \KwRet\;
 }{}

   \SetKwProg{myproca}{Procedure}{}{}
  \myproca{\slcost{$\sites, \ports$}}{
    $W$ $\gets$ zeros($|\sites|, |\ports|$)\;
    \For{$i \in \{1, \dots, |\sites|\}$}{
        \For{$j \in \{1, \dots, |\ports|\}$}{
            $W[i][j]$ $\gets$ $\lVert \sites[i] - \ports[j] \rVert$\;
        }
    }
  \nl \KwRet $W$\;}

   \SetKwProg{myprocb}{Procedure}{}{}
  \myprocb{\orcost{$\sites, \ports$}}{
    $W$ $\gets$ zeros($|\sites|, |\ports|$)\;
    \For{$i \in \{1, \dots, |\sites|\}$}{
        \For{$j \in \{1, \dots, |\ports|\}$}{
            %$W[i][j]$ $\gets$ Concentric arc from $\sites[i]$ to radial line through $O$ and $\ports[j]$ + radial line segment from concentric circle containing $\sites[i]$ to $\ports[j]$
            $W[i][j]$ $\gets$ Length of arc from $\sites[i]$ to $\ports[j]$\;
        }
    }
  \nl \KwRet $W$\;}
\end{algorithm}

The next step in our algorithm is establishing the length of every possible leader, considering every point feature and candidate port combination. The lengths of SL- and OR-leaders differ, as explained in \Cref{sec:design_space}, and those lengths are computed in procedures \texttt{computeCostSL} and \texttt{computeCostOR} of \Cref{alg:uniform}.

Once all distances between features and port candidates have been computed, we can use existing algorithms, such as Munkres' algorithm~\cite{munkres1957algorithms}, to match features to port candidates, such that the total leader length is minimized~\cite{fredman1987fibonacci}. As proven in~\cite{BonerathNTWW24}, the assignment can be obtained in $O(n^3)$ time and is both optimal and has the additional property that no leaders cross, regardless of which of the two leader styles is used.

% \begin{itemize}
%     \item Compute set of candidate label positions relative to anchor
%     \item Compute cost matrix of site/candidate pairs
%     \item Compute minimal cost matching
%     \item Connect each site to the assigned label candidate 
% \end{itemize}

\subsection{Non-uniform Labels}\label{sec:uniform_algo}

In contrast to the case for uniform labels, optimizing leader length for 
% an OR-labeling with 
non-uniform labels is \NP-hard~\cite{BonerathNTWW24} (for SL- and OR-labelings).
While the reduction presented in~\cite{BonerathNTWW24} does not assume the presence of an anchor $A$, the instance of the reduction is constructed in such a way that there is an optimal labeling in which a particular label will always be placed in the same location. Hence, the problem stays \NP-hard, even with the predefined $A$. 
\revOld{Moreover, we know for some instances that there is no crossing-free solution 
%(see \cref{sec:counterexample}).}
(see the counter example in the supplemental material).}
%\textcolor{red}{This is actually quite different now from the problem we considered (anchor point and only short leaders considered)}
Therefore, we first turn to a well-established optimization method for \NP-hard problems, namely mathematical programming \revOld{to establish an optimal baseline to compare the heuristics against}.

\subsubsection{Exact Algorithms}
We can formulate a mixed integer linear program (\textsf{MIP}) for computing optimal OR-labelings since crossings between OR-leaders can be determined combinatorially and the length of an OR-leader can be expressed as a linear function of the polar coordinates of its site and port.
However, for straight-line leaders, we have to compute the Euclidean length of a line segment to determine an optimal solution, \rev{which requires the use of the more expressive quadratic integer programming model (\textsf{QIP}).
%and the mathematical programming model to compute an SL-labeling is a quadratic integer program (\textsf{QIP}). 
For both models, the objective function is to find a label order of label that minimizes the total leader length.}

\newcommand{\portangle}{\beta}
\newcommand{\portdist}{d}

\mypar{The OR-model. }
%\textcolor{red}{Short version:}
We sketch the construction of the \textsf{MIP}.
%A full description can be found in Section~\ref{app:MIP}.
A full description can be found in the supplemental material.
The \textsf{MIP} uses a set of binary variables to encode a transitive order over all ports.
\rev{Each binary variable represents a pairwise ordering decision, indicating whether label $A$ appears before or after label $B$ in the counter-clockwise circular label order.}
Then, the angular component of the polar coordinate of a port is represented with a real-valued variable.
Since a labeling is completely characterized by the order of the labels, the port angle variables can be set based on the binary variables.
Next, the length of a leader in such labeling can be calculated based on the port angle variable and the angle of a point feature (as specified in the input).
The sum of all these angle differences is minimized to obtain a leader length-minimal labeling.
Finally, we can determine combinatorially if two leaders cross solely based on the angle variables of their ports. 
\rev{Intuitively, two leaders cross whenever the circular order of their labels is inconsistent with the circular order of their associated features.}
Hence, a crossing-free labeling can be encoded using binary indicator variables.

\mypar{The SL-model. }
%\textcolor{red}{Short version:}
%We again only sketch this \textsf{QIP} here and refer to Section~\ref{app:QIP} for a full description.
We again only sketch the \textsf{QIP} here and refer to the supplemental material for a full description.
The \textsf{QIP} shares similarities with the \textsf{MIP}.
The port positions are represented, and transitivity is enforced in the same way.
However, the length of a leader is determined using the Euclidean distance between a feature point and its port.
By adding the following (non-linear) constraints, we optimize the total leader length in the objective function.

\[ \text{Minimize: } \sum_{i=1}^{n} \gamma_i\]
\[ \forall 1\leq i \leq n: \gamma_i \geq \sqrt{(\site_i.\distance)^2 + R^2 - 2\cdot\site_i.\distance\cdot R \cos(\site_i.\anglec - \portangle_i)},\]
where the $\beta_i$ variable encodes the angle of a feature's port.
\rev{Moreover, the trigonometric functions and the square root in the constraint need to be approximated by piecewise linear functions. Unlike OR-leaders, crossings between straight-line leaders depend not only on the circular order of ports but also on their exact geometry, requiring an explicit geometric crossing test.}
%It remains to enforce a crossing-free solution.
This is done via careful case distinction.
For two leaders $p_1q_1$ and $p_2q_2$, consider the supporting lines $\ell_1$ and $\ell_2$, respectively.
We check if (i) $p_1$ and (ii) $q_1$ are on different sides of $\ell_2$ and if (iii) $p_2$ and (iv) $q_2$ are on different sides of $\ell_1$.
If and only if all four checks are true, the leaders cross.
This is again encoded using quadratic constraints and binary indicator variables.

\subsubsection{Heuristic Approaches}
Considering that the computation of a leader length minimal solution for non-uniform labels is already \NP-hard~\cite{BonerathNTWW24}, we also make use of a heuristic approach for this setting.
To obtain an initial solution, we obtain an order by pretending that all labels are, in fact, of uniform size and run \Cref{alg:uniform}.
%This order completely determines an initial order for non-uniform labels.
We proceed by computing initial label positions from this order. 
If this solution contains no crossings, we terminate.
Otherwise, we choose two arbitrary features whose leaders cross.
Then, we simply exchange these two labels in the order, yielding a new solution.
We repeat until the solution is crossing-free or until a previously fixed upper bound on the number of uncrossing operations is reached.

Note that there is no theoretical guarantee that this procedure terminates with a crossing-free solution on every instance. 
In the supplemental material, we present an example for which no feasible solution exists.
However, we have not encountered any instance during experimental evaluation for which this algorithm did not terminate successfully.
\section{Computational Experiments}
\label{sec:compExp}

While we know that our algorithms will compute leader-length optimal labeling in polynomial time for uniform label sizes, we do not have any theoretical guarantees for the heuristics. Hence, we conducted a computational experiment to evaluate the heuristics in terms of runtime and total leader length. Additionally, we also report runtime to compute exact solutions. Our experiment focuses on investigating the following questions:

\begin{itemize}[leftmargin=1em,labelwidth=*,align=left,  parsep=0.2em, itemsep=0.1em, topsep=0.2em]
    \item[\textbf{A.}] Is the runtime of the \textsf{MIP} or \textsf{QIP} practical in a general setting?
    \item[\textbf{B.}] What is the computational effort (runtime) of our heuristic?
    \item[\textbf{C.}] What is the difference in leader length between the exact and heuristic solutions?
\end{itemize}

\subsection{Experimental Setup}\label{sec:compExpSetup}

\mypar{Instances.} For our computational experiment, we decided to generate a synthetic dataset of instances that correspond to feature configurations that we expect to encounter in real-world instances. We used synthetic data over real-world data as it gives us the possibility to design instances that exhibit specific characteristics that we could control through parameters that potentially influence the solution quality of the heuristics. 

When generating an instance, we set the following parameters: \emph{instance size}, \emph{feature distribution}, and \emph{label size}. 
The instance size $n \in \{5, 6, \dots, 20\}$ represents the number of features and covers typical sizes that we would expect in a real-world scenario. 
The distribution of points $\{D_u, D_{u+o}, D_{o}\}$ determines the position of features. Generally, we assume that the radius of the boundary circle is $200$ units, and all generated points lie inside a radius of $150$ units. 
Furthermore, we ensure that all features are at least $5$ units apart \rev{to allow discernible leader lines and avoid numerical precision errors}. 
% The choice of radius reflects approximately the available pixels on a large (213 dpi) smartwatch screen centered over a zoomed-out area of interest. 
The minimum distance between features guarantees a non-overlapping layout. 
Now, $D_u$ represents instances that exhibit a uniform distribution of features. 
We determine each feature's position in polar coordinates by assigning an angle $\theta=[0, 2 \pi]$ and radius $r = [0, 150]$ uniformly at random. 
Such instances reflect a balanced distribution of point features, e.g., a selection of POIs around the position of the user. 
$D_o$ represents instances where all features are off-center. 
First, we determined a center point for the off-center distribution by picking an angle $\theta=[0, 2 \pi]$ and radius $r = [0, 100]$ uniformly at random. 
This center point is then used to offset the position for all features. 
For each feature, we select an angle $\theta=[0, 2 \pi]$ uniformly at random and select a radius from a normal distribution with standard deviation $\sigma = 75$. 
If a feature would be placed outside the radius of $150$ units, we discard the position and compute a new position. Such instances could, e.g., represent that the user is away from a small city containing all POIs. 
Lastly, instances with $D_{u+o}$ represent instances where half the features are concentrated around a specific point while the other half is uniformly distributed (e.g. inside a city with some POIs around, but most POIs are concentrated around a specific location like the old town). 
We combine the aforementioned procedures to achieve such an instance.
% Such instances could represent that the user is inside a city with some POIs around, but most POIs are concentrated around a specific location, e.g., the old town.
Finally, the label size of a feature is determined by selecting uniformly at random from the range $[1,5]$ (label lengths are normalized by the total label length to match the circumference). %For example, this could mean we allow labels to have between 4 and 20 characters.
With the specified parameters, we generated five instances for each combination, i.e., $16 \times \, 3 \times \, 5 = 240$ instances.

\mypar{Computational environment.}
All experiments were performed on a compute cluster. Each node is equipped with two AMD EPYC 7402, 2.80 GHz, 24-core processors and 1 TB of RAM. Each computation of an instance was scheduled as a single job on the cluster, with a maximum runtime of four hours and a maximum RAM allocation of 64GB. \revOld{For the heuristics, we only allowed single-core allocation.} All implementations were done in Python 3.12. The \textsf{MIP} and \textsf{QIP} formulations given in \Cref{sec:algorithms} were optimized using the Gurobi\footnote{\url{https://www.gurobi.com/}} optimizer (Version 11.0). 
\revOld{
Although we conducted the experiments on a high-performance cluster to ensure reproducible timing measurements, the computational demands of the heuristic algorithms are modest. 
For instance, all instances with up to $n=20$ labels are solved within milliseconds in case of the heuristics. 
For example, modern smartwatches use dual-core processors with around 1.0GHz~\cite{ShaheenAAA24} clock speed.
Even under a conservative estimate, scaling solely by clock frequency and ignoring architectural differences, this would increase the runtime by only a factor of about three compared to our cluster.
Thus, while the reported runtimes were obtained on a powerful machine, the same computations can easily be performed on embedded or mobile hardware.
}
%We chose powerful hardware over something comparable to a mobile device as we expected our models to be computationally expensive.

\mypar{Metrics.}
For each instance in our dataset, we computed a labeling with SL- and OR-leaders using exact and heuristic algorithms. 
We repeatedly ran each algorithm five times and reported the median wall clock time for computing a labeling.
To compare solution quality, we store the total leader length $\TLL$ for each labeling.
The relative total leader length $\TLL_r = \TLL / \TLL^*$ is the ratio of total leader length $\TLL$ of a heuristic solution to the corresponding optimal solution $\TLL^*$.

\subsection{Results}

Here we discuss the results from the computational experiment. This section follows the earlier stated research questions. Additional plots and the evaluation code are in a Python notebook in the supplemental material on \href{https://osf.io/3ekjb/?view_only=b8f3cde66c2543439897834d4b463b13}{OSF}. 
For all statistical tests, we assume a significance level of $\alpha = 0.05$. 
For effect size, we report the difference of means and bootstrapped~\cite{efron1986bootstrap} confidence intervals.

%For effect size we report the Common Language Effect Size ($\cles$), i.e., the probability that a sample from distribution A will be greater than a sample from distribution B.

\mypar{Runtime of the \textsf{IP} models (Question A).} 
The first part of the experimental evaluation focuses on computing a labeling with the \textsf{MIP} and \textsf{QIP} models in practice. 
For OR-leaders, we were able to solve all instances to optimality under the given constraints. 
However, we could only solve $88.0\%$ of all instances for SL-leaders.
% Moreover, we performed a Kruskal-Wallis test between $D_u$, $D_{u+o}$ and $D_o$ that showed significant differences ($p < 0.01$). 
% We followed up with a pairwise Mann-Whitney U test, which revealed there is a significant difference ($p < 0.01$) between all feature distributions. 
% Further, the effect size revealed that we can deduce an order ($D_u < D_{u+o} < D_{o}$) if we are able to solve an instance.  
% Hence, we can conclude that feature distribution has an impact on the solvability of instances -- the less uniform features are distributed, the less likely it is to get a result at all.

Next, we plotted the runtime against instance size, which is presented in the supplemental material. 
For OR-leaders, we were able to compute small instances ($n \le 12$) in under a second. 
The runtime rises exponentially, and computing the largest instances ($n = 20$) took approximately 100 seconds.   
The runtime behavior for computing a labeling with SL-leaders is worse. Even the smallest instances ($n = 5$) took 10 seconds, and successfully computing larger instances can take up to our time limit of four hours. We explain this behavior with the more complex quadratic formulation SL-leaders.
Lastly, we performed a Kruskal-Wallis test comparing $D_u$, $D_{u+o}$, and $D_o$, which showed no significant differences. 

To conclude, using \textsf{QIP} models to solve instances with SL-leaders has no practical relevance for now. Computing the labeling with OR-leaders for instances with a reasonable number of features could potentially be optimized further. 
However, for both leader types, the possibility of computing an optimal solution allows for a comparison of the performance of heuristics.

\mypar{Computational effort of the heuristics (Question B).}
Although we do not have a theoretical argument that our heuristics guarantee a crossing-free solution and termination, we can report that all instances in our experimental dataset terminated successfully. 
The mean wall clock runtime for computing the largest instance ($n = 20$) remained below 20ms for SL- and OR-leaders. Thus, we created more instances for up to 100 features. 
The runtime plot for the extended dataset, which can be found in the supplemental material, revealed that wall clock runtime remains below 100ms for up to 40 features. 
However, the runtime starts to exceed one second for $n > 90$.
Plotting the runtime of both leader types against the number of features reveals a slight difference between SL- and OR-leaders, which we confirmed with a Wilcoxon signed-rank test ($p < 0.01$). Computing OR-leaders (mean $407.02$ms, $[369.41, 446.8]$) is faster than computing SL-leaders (mean $753.08$ms, $[678.53, 838.79]$).

To conclude, computing a labeling with SL- and OR-leaders for $n \le 20$ is near-instant.
\rev{As argued earlier, mobile hardware is powerful enough to handle computation of a labeling, and interactive applications can recompute labelings with low latency.}

\begin{figure}[t]
    \centering
    \includegraphics[width=\linewidth]{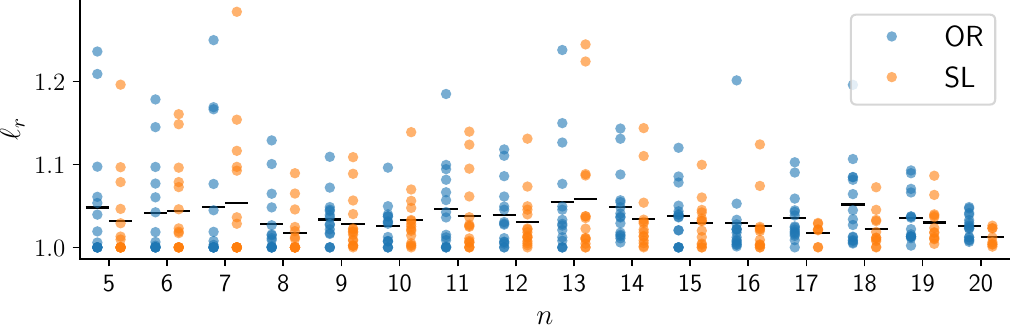}
    \caption{Relative total leader length $\TLL_r$ of OR-  and SL-leaders plotted against the number $n$ of features. Black bars indicate the mean.}
    % $\TLL_r$ of SL- and OR-leaders.
    \label{fig:solution_quality}
\end{figure}

\begin{figure*}[th]
  \centering
      \begin{subfigure}{0.25\linewidth}
      	\centering
      	\includegraphics[width=\linewidth]{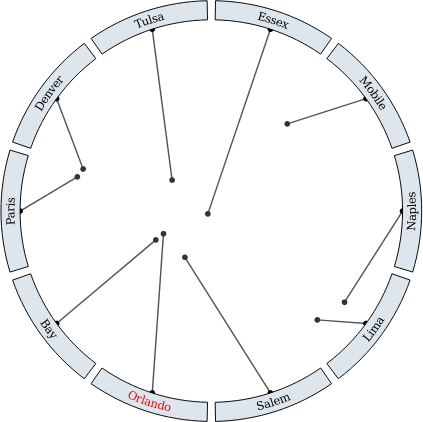}
      	\caption{$D_u$ -- uniform feature distribution}
      	\label{fig:study_3}
      \end{subfigure}
      \hfill
      \begin{subfigure}{0.25\linewidth}
      	\centering
      	\includegraphics[width=\linewidth]{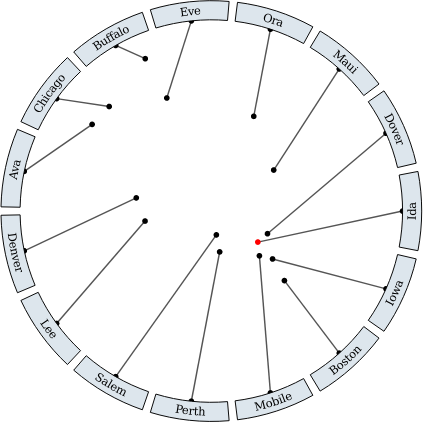}
      	%\caption{$D_{u+o}$ -- uniform and offcenter feature distribution}
        \caption{$D_{u+o}$ -- uniform + offcenter dist.}
      	\label{fig:study_2}
      \end{subfigure}%   
      \hfill
      \begin{subfigure}{0.25\linewidth}
      	\centering
      	\includegraphics[width=\linewidth]{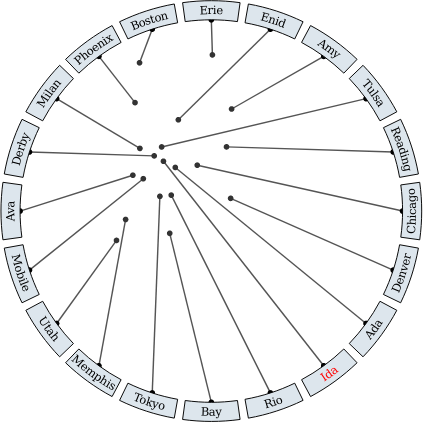}
      	\caption{$D_o$ -- offcenter feature distribution}
      	\label{fig:study_1}
      \end{subfigure}%   
  \caption{Stimuli of three instances with SL-leaders and uniform label length. (a) shows an uniform distribution ($D_u$) of 10 features where a target label ($T_L$) is given in red. (b) shows a uniform and off-center distribution ($D_{u+o}$) of 15 features where a target feature (task $T_\sites$) given in red. Lastly, (c) shows an off-center distribution ($D_u$) of 20 features where a target label (task $T_L$) is given in red.}

  \label{fig:stimuli}
\end{figure*}

\mypar{Total leader length of the heuristics (Question C).}
Lastly, we look at the solution quality in terms of the total leader length of our heuristics compared to the exact solutions. 
We only consider instances where an exact solution could be computed.
\rev{We provide a visual comparison on several instances and additional analysis in the supplemental material.}
Firstly, \Cref{fig:solution_quality} shows the relative total leader length of SL- and OR-leaders. 
The $\TLL_r$ for both leader styles ranges from $1.00$ to  $1.28$ with the mean value of OR-leaders ($\TLL_r = 1.033$, $[1.027, 1.039]$) being slightly better than SL-leaders ($\TLL_r = 1.040$, $[1.034, 1.047]$). 
To see the impact of instance size, we ran a Spearman correlation test which revealed that there is an indication ($p = 0.04)$ for OR-leaders ($r = 0.14$) but not for SL-leaders that instance size correlates with $\TLL_r$.
Lastly, we also tested the impact of feature distribution on the solution quality.
Here, we can report that a Kruskal-Wallis H-Test revealed no significant differences between $D_u$, $D_{o+u}$, and~$D_o$. 

To conclude, solution quality in terms of relative total leader length suggests that our heuristics are close to the optimal solution. 
\rev{As illustrated in \Cref{fig:solution_quality}, with increasing $n$, the heuristic solutions are already near-optimal, likely because the instances admit only few feasible crossing-free label orders, leaving little freedom for improvement beyond the initial solution.} However, we expected a higher impact of instance size or point distribution on $\TLL_r$. Unfortunately, due to the high computing time, we cannot generalize our result to instances with more feature points.

\section{User Experiment}
\label{sec:userExp}

A user study by Barth et al.~\cite{BarthGNN19} has previously investigated one-sided boundary labeling. We argue that our setting is distinct enough to warrant a separate investigation that focuses on the readability of orbital labeling with SL- and OR-leaders. 

\subsection{Participants and Setting}

We opted to conduct the study online, sacrificing the controlled environment of a lab study, in order to better access a wider audience. We designed the study as a \emph{within-group} experiment where we confronted each participant with both leader styles on 32 instances. For half of the instances, we used non-uniform label sizes. The target time for completing the study was between 10-20 minutes. Participants in the experiment were required to be at least 18 years of age and to have access to a screen with a sufficiently wide display and an appropriate input device. \rev{The rational behind not conducting the study on a small screen, for example a smartwatch, are the confounding factors such as movement, viewing distance, or input precision.} We considered screen sizes of desktop PCs, laptops, and tablets to be sufficient. Furthermore, we considered a computer mouse, trackpad, or pen, as appropriate, for input devices in the case of tablets. \rev{All participants provided informed consent prior to the study. Participation was voluntary, and participants were informed about the study procedure, the type of data collected, and their right to withdraw at any time without penalty. During the study, no institution involved required ethics board approval.}
%Additionally, we asked participants to switch to fullscreen mode of the browser and self-exclude if they had problems performing the tasks during the tutorial. 

We recruited participants via mailing lists and social media, and in total, we collected 54 complete responses. From the demography questions, we analyzed that of our participants $35$ identified as male, $18$ identified as female, and one as other. 32 participants reported to be between the ages of 25-34, with a near-equal split of the remaining participants over other age brackets. $44$ of our participants reported having at least a Bachelor's degree. 37 participants completed the study using a mouse, and 17 used a trackpad. 
%Almost all participants worked daily with a computer and $33$ reported that they play video games.
%Almost all participants reported that they had used map navigation on a mobile device before.

\subsection{Hypotheses}\label{sec:hypothesis}

For the user experiment, we formulated two alternative hypotheses. 
%If we state that labeling A outperforms labeling B on task accuracy, this means that labeling A has a higher accuracy. Similarly, if we state that labeling A outperforms labeling B on response time, then labeling A has a lower response time.

\begin{itemize}[leftmargin=1em,labelwidth=*,align=left,  parsep=0.2em, itemsep=0.05em, topsep=0.2em]
    \item \textbf{H1}: Task accuracy will have no significant difference between OR-leaders and SL-leaders.
    \item \textbf{H2}: Response time will have a significant difference between OR-leaders and SL-leaders. Participants will perform tasks faster with SL-leaders than OR-leaders.
\end{itemize}

The hypotheses are based on the findings of Barth et al.~\cite{BarthGNN19}, who showed that SL-leaders performed well in terms of response time compared to other leader types for one-sided boundary labeling. They could not identify significant differences regarding task accuracy between some leader types. We assume that their findings generalize to some extent to orbital labeling.

\subsection{Tasks}

We selected the same fundamental tasks as Barth et al.~\cite{BarthGNN19}.  

\begin{itemize}[leftmargin=1em,labelwidth=*,align=left,  parsep=0.2em, itemsep=0.05em, topsep=0.2em]
    \item \textbf{$T_\sites$:} Given a feature, find the associated label.
    \item \textbf{$T_L$:} Given a label, find the associated feature.
\end{itemize}

%The reason for selecting these two abstract tasks is that our investigation focused on the performance differences between SL- and OR-leaders. 
\rev{
We selected two tasks that align with the two hypotheses and required participants to follow the leader from feature to label, or vice versa, by clicking either the label or the feature.}
While the label is the larger target, it should have no impact on the results, as we did not intend to compare between the tasks.
%Also, both tasks allow us to visually highlight the feature or label, reducing confounding factors of visually searching for the asked feature or label first. 

\subsection{Datasets}

We used a similar approach as described in \Cref{sec:compExpSetup}. We decided on three instance sizes $n \in \{10, 15, 20\}$ and generated for each task in $\{T_\sites, T_L\}$ an instance with $D_u$, $D_o$, and $D_{o+u}$ feature distribution. 
Additionally, we generated a second dataset for non-uniform label sizes, where we assigned each feature a label with size in $[1,5]$ uniformly at random. 
Furthermore, we augmented each instance with random city names that correspond in character count to the label length. 
We reason that this is a more realistic scenario than omitting label text or using a single character. 
Lastly, we hand-picked a feature for each dataset that would later serve as the target of the experiment, i.e., we provided a target feature, and the participant had to find the corresponding label or vice versa. 
We aimed to balance the task difficulty by carefully selecting features: avoiding those that are isolated from other features to prevent the task from being too easy, and avoiding options like OR-leaders that are too close to avoid making it excessively challenging.

\subsection{Stimuli}

\begin{figure*}[t]
    \centering
    \includegraphics[width=\textwidth]{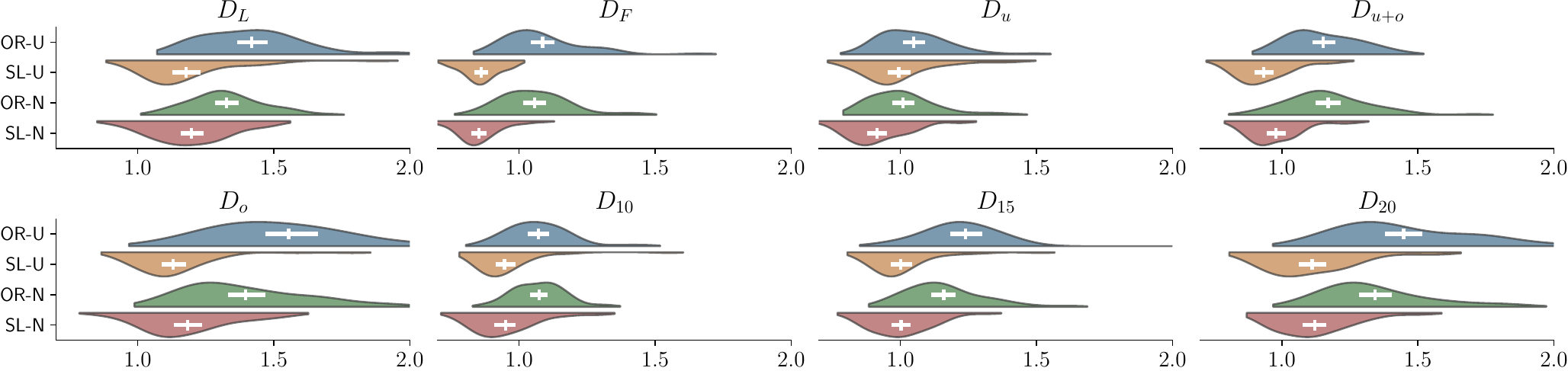}
    \caption{Participants' response time for each stimulus partition and both leader types with uniform (U) and non-uniform (N) label length. Inside the violin plots a dot plot indicates the mean and 95\% confidence interval. \rev{All instances are partitioned into their characteristics such as number of labels ($D_{10}$, $D_{15}$, $D_{20}$), feature distribution (uniform $D_{u}$, uniform and non-uniform $D_{u+o}$, non-uniform $D_{o}$), and wether label ($D_{L}$) or feature ($D_{F}$) was the target.}}
    \label{fig:response_time} 
\end{figure*}

We used our own implementation to compute the visualizations that are later shown as stimuli to participants. 
Examples can be seen in \Cref{fig:stimuli}. 
We used the \textsf{MIP} and \textsf{QIP} models to compute optimal solutions for non-uniform label size instances.
Hence, we avoid that our study results are dependent on suboptimal heuristic solutions.

We rendered all instances as an SVG. The circular boundary has a diameter of 440px with labels of 20px height, resulting in a total diameter of 480px. 
On a typical computer screen, under the assumption that participants sit at a sufficient distance away, this should result in a visualization that falls within central vision~\cite{loschkySBYB}. 
%We argue that this is similar to looking at a smartwatch screen from a typically shorter distance. 

\revOld{To isolate the effects of leader geometry from other visual factors, we chose a plain white background instead of embedding the features in a map. Although maps represent a typical application scenario, they introduce visual noise, spatial cues, and semantic expectations that can mask or override the specific effects of leader type on performance. Our goal was to assess the perceptual and cognitive differences between SL- and OR-leaders in a controlled environment, free from such confounding influences.} The labels were filled with a light-blue background to differentiate them better. The text was rendered in 10pt font size. Features were rendered as dark grey disks with a diameter of 5 pixels. The leader lines were rendered as dark grey lines. If a label or feature was the target of an instance, we rendered either the label text or disk in red to highlight the visual element. We expect that highlighting reduces the confounding factor of a visual search. We added hover interactivity to a stimulus by highlighting the counterpart of the target. For example, if we asked a participant to find the label of a target feature, we would highlight the label in red if the participant hovered over it. This should better reflect the participant's intent during the experiment. Lastly, as we generate two stimuli (SL- and OR-leaders) from one instance, we counteracted learning effects by mirroring the coordinates of features on the x-axis for one of the stimuli.

\subsection{Experimental Procedure}

In total, each participant had to perform $2\;\text{label conditions} \;\times 2\,\text{leader conditions} \;\times 2\,\text{tasks}\;\times 9\,\text{instances} =72\, \text{trials}$. 
The experiment followed a five-stage template of (1) consent and screening, (2) tutorial, (3) formal study, (4) post-task questionnaire, and (5) demographic questions. 

In stage (1), participants were informed about the study's purpose, procedure, requirements, data policy, and consent form. Upon consenting, they enabled fullscreen mode in their browser, self-excluded if device limitations arose, and we logged browser agent details. We introduced map labeling and external labeling concepts and demonstrated SL- and OR-leaders using a real-world example. Stage (2) involved two tutorial blocks. In the first block, participants learned the tasks, practiced interacting with the visualization, and solved tasks using both leader styles on uniform label sizes. The second block introduced non-uniform label sizes using a similar procedure. Participants had to correctly complete all tutorial tasks before proceeding.
In stage (3), participants began the formal study and were instructed to solve tasks ``correctly while maintaining some speed''. To minimize learning effects, the order of blocks and questions was randomized for each participant. Stimuli appeared only after clicking a ``Start Task'' button, centrally positioned to neutralize mouse location. After the trials, participants completed a post-task questionnaire (4), rating preference, aesthetics, and confidence using a 5-point Likert scale. Finally, stage (5) collected demographic information.

\subsection{Pilot}

We conducted a pilot study with three non-expert participants and two visualization researchers experienced in user study design. Each participant was interviewed about their experience, and the experiment was adapted based on the findings. Participants reported no concerns about the study's design, flow, difficulty, or duration. The pilot study initially included 64 stimuli, differing from the main experiment in feature count and distribution. The pilot dataset used 8 and 16 features with $D_u$ and $D_o$ distributions, creating 32 stimuli. An additional 32 stimuli, with identical parameters but a hard time limit, were included to simulate stress, akin to a brief glance at a smartwatch. However, participants indicated they did not perceive a difference between tasks with and without a time limit. As a result, the time limit was removed to avoid complicating the setup. In post-study interviews, participants noted that tasks with the 8-feature dataset were often solvable without following the leader line. Consequently, we expanded to ${10,15,20}$ features to investigate finer effects and introduced the $D_{u+o}$ distribution.
%Lastly, one participant vocalized concern that the mouse pointer position after clicking ``Start Task'' could potentially highlight a non-target feature.  

\subsection{Results and Analysis}

To better analyze the differences of the stimuli, we create several partitions over them. First, we partition the stimuli by task into $D_{\sites}$ and $D_{L}$, i.e., all tasks where the participant had either a label or feature given. Second, we partition the stimuli by size into $D_{10}$, $D_{15}$ and $D_{20}$. Third, we partition the stimuli by feature distribution into $D_u$, $D_o$, and $D_{u+o}$. Each partition represents a \emph{class} of stimuli.
During the analysis, we used the mean task accuracy and response time for each participant's stimuli in a class. 

Below, we describe the statistical tests for task accuracy, response time, and qualitative feedback. Generally, we performed a post-hoc analysis if a statistical significance ($\alpha = 0.05$) was given. We used a Wilcoxon signed-rank test with splitting ties~\cite{pratt1959remarks} to compare between leader types. When comparing classes for the same leader type, we used a Friedman test and a post-hoc Mann-Whitney U test with one-step Bonferroni correction. For the effect size, we report the difference of means with bootstrapped $95\%$ confidence intervals. 
All plots and the statistical evaluation can be found in the supplemental material on \href{https://osf.io/3ekjb/?view_only=b8f3cde66c2543439897834d4b463b13}{OSF}. 

\mypar{Task accuracy.}
The plot of task accuracy according to partition and leader type is provided in the supplemental material.
For uniform labels, we did not find statistically significant differences between SL- and OR-leaders \textit{within classes} besides $D_{15}$ ($p = 0.049$) where SL-leaders slightly outperform OR-leaders ($0.03$, $[0.01, 0.06]$).
Overall, all participants had a mean greater than $0.95$ with the median being $1.0$.
For non-uniform labels, we did not find any indications of significant differences (median $> 0.98$). 

Next, we investigated differences \textit{between classes} within SL- and OR-leaders for uniform labels. 
There is a significant difference between $D_u$, $D_o$ and $D_{u+o}$ for SL-leaders ($p = 0.01$) but not for OR-leaders. The post-hoc test showed a significant difference between $D_{o}$ with $D_{u+o}$ ($0.03$, $[0.01, 0.05]$) and $D_u$ ($0.03$, $[0.01, 0.05]$). Instance size had a significant difference for SL- and OR-leaders. The post-hoc test showed for OR-leaders that participants had higher accuracy with $D_{10}$ compared to $D_{15}$ ($0.03$, $[0.01, 0.06]$). For SL-leaders participants had lower accuracy with $D_{20}$ compared to $D_{10}$ ($0.03$, $[0.01, 0.05]$) and $D_{15}$ ($0.03$, $[0.01, 0.05]$). 

For \textit{non-uniform labels}, we only found one difference \textit{between classes}. For OR-leaders participants had lower accuracy with $D_o$ compared to with $D_{u+o}$ ($0.015$, $[0.00, 0.03]$) and $D_u$ ($0.015$, $[0.00, 0.03]$).
For uniform and non-uniform labels, we did not find significant differences between tasks.

To conclude, even when we found significant differences, the effect size remained small, confirming hypothesis H1.  
\rev{This is in line with the experiment of Barth et al.~\cite{BarthGNN19} comparing different leader types in rectangular boundary labeling.}
%Furthermore, our analysis suggests that feature size has a bigger impact than feature distribution for SL-leaders and OR-leaders. OR-leaders seem to also be more impacted by feature distribution.

\mypar{Response time.}
For each participant, we normalized the response time by the respective median response time over all stimuli to account for reaction time.
A plot of the normalized response time for each class is shown in \Cref{fig:response_time}.
Here, we report the difference of the mean and confidence interval in seconds.
We first looked at the differences between SL- and OR-leaders \textit{within classes} for \textit{uniform} label sizes.
There are statistically significant differences ($p < 0.01$) indicating that SL-leaders outperform OR-leaders across all classes.  
Regarding feature distribution we noticed the least strong effect size in $D_u$ ($0.05$, $[0.01, 0.09]$), followed by $D_{u+0}$ ($0.22$, $[0.18, 0.26]$) and $D_o$ ($0.42$, $[0.35, 0.51]$) suggesting that SL-leaders handle clustered features better. We made a similar observation regarding  $D_{10}$ ($0.12$, $[0.07, 0.16]$), $D_{15}$ ($0.23$, $[0.19, 0.31]$), and $D_{20}$ ($0.33$, $[0.26, 0.40]$) giving a slight indication that SL-leaders perform better with an increasing number of features. Lastly, $D_{\sites}$ ($0.22$, $[0.19, 0.26]$) and $D_{L}$ ($0.24$, $[0.19, 0.30]$) differ. 

We investigated differences \textit{between classes} within SL- and OR-leaders for \textit{uniform} labels. 
There is a significant difference between $D_u$, $D_o$, and $D_{u+o}$ for both leader styles ($p = 0.01$). For both leader types, the pairwise difference implies a strict order of $D_u < D_{u+o} < D_o$. We observed a similar effect for instance size with $D_{10} < D_{15} < D_{20}$, suggesting increased difficulty. Effect sizes are in the supplemental material.
\rev{Note that for SL- and OR-leaders, participants performed worse for $D_u$ and $D_{u+o}$. A potential reason is that instances for $D_{u+o}$ were easier to solve.}

For \textit{non-uniform} label sizes, we report significant differences ($p < 0.01$) between SL- and OR-leaders for all classes. 
Regarding feature distribution we noticed the least strong effect size in $D_u$ ($0.09$, $[0.05, 0.14]$), followed by $D_{u+0}$ ($0.19$, $[0.15, 0.24]$) and $D_o$ ($0.21$, $[0.16, 0.27]$) suggesting that SL-leaders handle concentrated features better. We made a similar observation regarding  $D_{10}$ ($0.12$, $[0.09, 0.16]$), $D_{15}$ ($0.15$, $[0.11, 0.20]$), and $D_{20}$ ($0.22$, $[0.17, 0.28]$) giving a slight indication that SL-leaders perform better with an increasing number of features. Lastly, $D_{\sites}$ ($0.20$, $[0.17, 0.24]$) and $D_{L}$ ($0.13$, $[0.08, 0.18]$) differ. Regarding differences within leader type but \textit{between} classes, we made similar observations to uniform label sizes. This can also be found in the supplemental material.

To conclude, we found strong evidence that response time is lower for SL-leaders than for OR-leaders regardless of label size, number of labels, and distribution. 
%\rev{Additionally, the smoother distribution of OR-leaders in contrast to SL-leaders suggests that participants likely take more time following the leader, rather than jumping to the target.} 
This confirms our hypothesis H2. As expected, instance size and feature distribution impact response time. To our surprise, we also found differences between tasks that might have been caused by our experimental design.
\rev{In contrast to Barth et al.~\cite{BarthGNN19}, SL-leaders outperform OR-leaders on task completion time, while their experiment showed that parallel-orthogonal leaders are significantly faster than SL-leaders for rectangular boundary labeling.}
%indicating that the reading direction of a leader influences the response time to some extent.

\begin{figure}
    \centering
    % \begin{subfigure}{\linewidth}
    %     \includegraphics[width=\linewidth]{figures/Preference.pdf}
    %     \caption{Preference}
    %     \label{fig:Preference}
    % \end{subfigure}
    \begin{subfigure}{0.95\linewidth}
        \includegraphics[width=\linewidth]{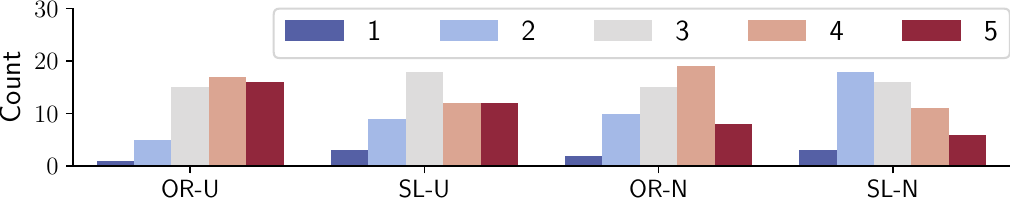}
        \caption{Aesthetics}
        \label{fig:Aesthetics}
    \end{subfigure}
    \begin{subfigure}{0.95\linewidth}
        \includegraphics[width=\linewidth]{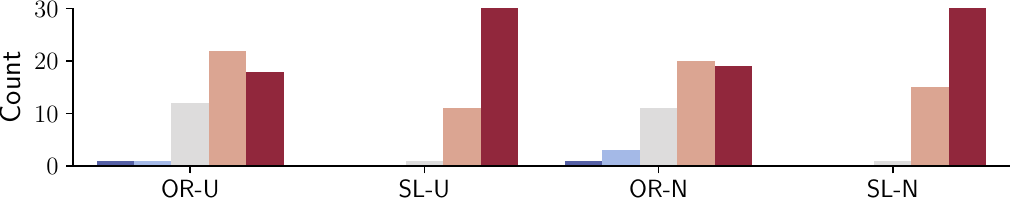}
        \caption{Confidence}
        \label{fig:Confidence}
    \end{subfigure}
    \caption{Qualitative answers of the participants for both leader types with uniform (U) and non-uniform (NU) label lengths.}
    \label{fig:qualitative}
\end{figure}

\mypar{Qualitative feedback.}
We asked participants to report their experience on a five-point Likert scale (1 = worst, 5 = best).
The plot of the statistically significant results can be seen in \Cref{fig:qualitative}.
For confidence, we found statistically significant ($p < 0.01$) differences. The post-hoc test showed that for uniform ($0.73$, $[0.54, 0.96]$) and non-uniform ($0.69$, $[0.46, 0.90]$) labels, participants consistently preferred SL-leaders over OR-leaders. We did not find pairwise differences for the same leaders between uniform and non-uniform labels.
Answers regarding aesthetics also differ significantly ($p < 0.01$). The post-hoc test showed that participants preferred OR-leaders over SL-leader for uniform ($0.34$, $[0.08, 0.76]$) and non-uniform ($0.40$, $[0.02, 0.79]$) labels. Furthermore, participants found OR- ($0.36$, $[0.11, 0.54]$) and SL-leaders ($0.42$, $[0.19, 0.67]$) more aesthetic when combined with uniform labels.
We did not find any significant difference regarding participants' preferences.
Lastly, we did not find any correlation between aesthetics, preference, or confidence with accuracy or response time.

To conclude, the statistical tests showed that participants seem to find OR-leaders more aesthetically pleasing but feel more confident with SL-leaders. There is no indication that participants prefer one over the other. This notion was also reflected in the free-form text field. A recurring theme was vocalized by a participant as 
%``Following straight line leaders was most of the times easy, for orbital-radial leaders I found following the leaders in some settings more difficult'', or 
``Both lines are easy to follow. The orbital-radial leader lines become more difficult, the moment you put them directly next to each other''. 
\rev{In Barth et al.~\cite{BarthGNN19}, participants preferred parallel-orthogonal leaders over SL-leaders, which is confirmed by our experiment.}

\section{Discussion}
\label{sec:discussion}

Lastly, we want to summarize our findings, discuss limitations, and outline potential future work of orbital labeling.

% Orbital labeling was introduced only recently on a theoretical level. In this paper, we transfer this concept to a visualization context by identifying the variants that are most relevant for circular interfaces or radial visualization designs, and constraining others to enable efficient heuristic computation. 
% This step transforms the concept from a purely theoretical construct into a practical labeling technique.

\revOld{
Orbital boundary labeling was introduced only recently on a theoretical level~\cite{BonerathNTWW24}. 
That work defined a broad family of variants differing in port placement, leader geometry, and label ordering, many of which are of mainly theoretical interest. 
In this paper, we apply the concept to a visualization context, focusing on the variants most relevant to circular interfaces and radial visualization designs, namely, fixed-port orbital-radial and straight-line leaders, for both uniform and non-uniform label sizes. 
Variants that depend on free port positioning or ordering constraints are omitted, as they increase complexity or are too restrictive for visualization purposes. 
We also introduce a practical simplification by aligning the first label to a fixed reference angle, which enables efficient heuristic computation while preserving layout generality. 
Together, these adaptations transform orbital labeling from a purely theoretical construct into a practical and implementable labeling technique suitable for real visual interfaces.
\rev{However, for some variants, we trade generality for computational tractability.}
}

\begin{figure}[t]
    \centering
    \begin{subfigure}{.47\linewidth}
        \centering
        \includegraphics[width=\linewidth]{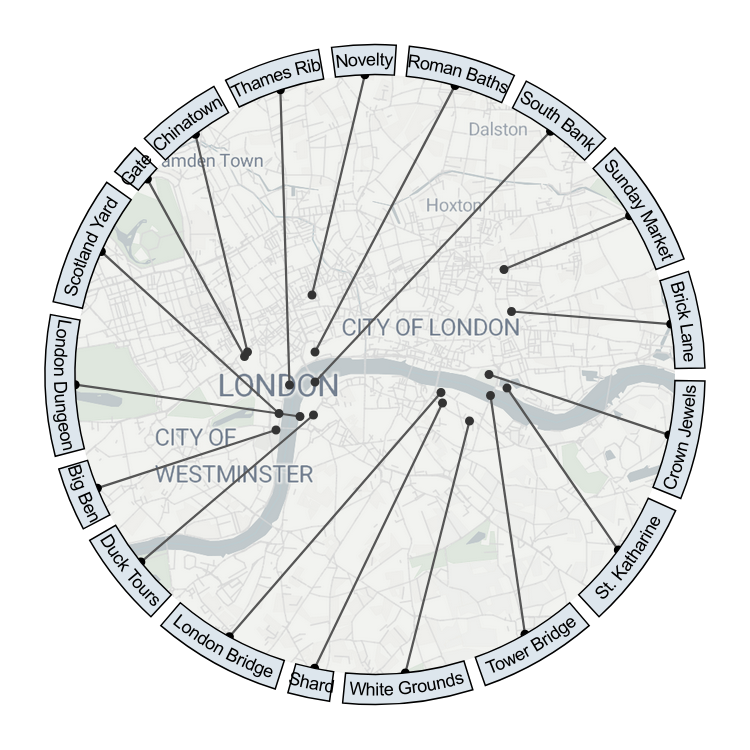}
        \subcaption{SL-leader}
    \end{subfigure}
    \hfill
    \begin{subfigure}{.47\linewidth}
        \centering
        \includegraphics[width=\linewidth]{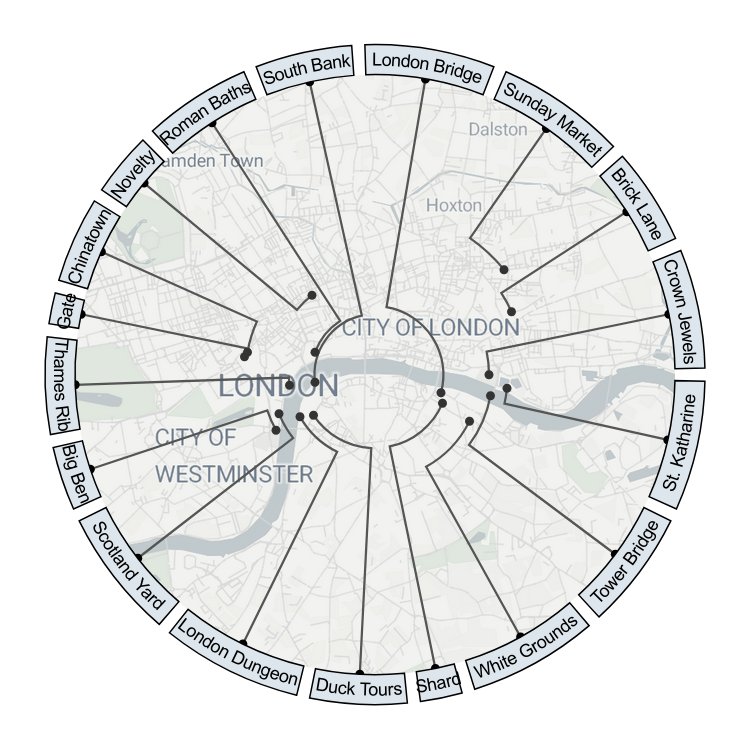}
        \subcaption{OR-leader}
    \end{subfigure}
    \caption{A real-world example showing orbital boundary labeling for non-uniform labels for 18 \del{tourist} attractions in London. The attractions were selected randomly by an automated query to OpenStreetMap.}
    \label{fig:RL_example}
\end{figure}

\revOld{
We implement one existing algorithm for uniform label sizes~\cite{BonerathNTWW24}, but we also present novel algorithms for both exact (\textsf{MIP}, \textsf{QIP}) and heuristic solutions for minimizing total leader length for non-uniform labels.}
The computational experiment demonstrates that our heuristics \del{for computing an orbital labeling} always yield a crossing-free solution, for which the total leader length differs from the optimal solution by at most a factor of 1.28. 
Our heuristics are easy to implement and fast.
Although experiments were executed on a desktop workstation, the measured runtimes translate directly to embedded devices because our heuristics operate in milliseconds for realistic instance sizes.
%This makes orbital labeling viable to incorporate in interactive applications and devices with limited computational power. 
The runtime of the QIP models is too slow to have practical relevance, but they provide a ground-truth for developing heuristics.    

\revOld{
We argue that orbital labeling can be used effectively in real circular interfaces. 
Typical small screens, such as smartwatches, can comfortably display around 10-16 short labels along the annular orbit without compromising legibility, while larger circular or radial visualization designs are not limited by this constraint. 
\rev{Due to screen size, smartwatches often have stricter minimum font size requirements, further limiting the number of labels.}
\Cref{fig:RL_example} shows a real-world example with 18 labels, which we think is still feasible.
Beyond smartwatches, potential applications include automotive dashboards, cockpit dials, circular widgets, and radial visualization designs, where external labeling helps preserve the central display area and maintain a clean visual layout.
\rev{More varied examples can be found in the supplemental material.}
}

Our user study reveals that SL- and OR-leaders perform equally well in terms of accuracy; however, SL-leaders enable significantly faster task completion time and higher confidence, while OR-leaders are perceived as slightly more aesthetically pleasing. 
We therefore recommend SL-leaders when response time is critical and OR-leaders when visual balance is prioritized. 
Participants noted that OR-leaders can become harder to read when orbital segments are close; this motivates exploring variants such as \emph{ROR-leaders}, which add a radial segment to increase separation between orbital arcs, as a promising direction for future research.

\mypar{Limitations.}
%The implemented algorithms make some assumptions which restrict the problem. The existence of the anchor $A$ could be relaxed and the implications of free rotation have been investigated~\cite{BonerathNTWW24}. 
    %Further the assumptions that any OR-leader will always use the shorter circular arc is an interesting open theoretical question.
    %makes valid crossing-free solutions impossible. However, we have not encountered any instances, where a leader length optimal solution requires the use of the longer circular arc. If such an instance exists or not is an interesting open theoretical question.
Our computational experiment used a synthetic dataset. We tried to model several potential real-world scenarios, but we did not investigate how well these generalize. 
\rev{While the relevant variants of orbital labeling were reasonably selected, we did not conduct a design study to empirically confirm the selection.}
Furthermore, our study was conducted online. Hence, we had no control over how participants perceived the stimuli. Our participants were predominantly well-educated and young; thus, it is unclear how well our findings on orbital labeling generalize to other user groups. Furthermore, a map or spatial geometry in the background can influence the perception of the labeling. 
\rev{Another factor that can influence readability is the orientation of labels, as vertical labels are potentially more difficult to read.} 
\rev{We expect that some of our findings may generalize to smartwatches; however, this remains to be validated in a dedicated study.} Lastly, we only showed a limited number of stimuli to each participant. Thus, we are not sure how well factors such as instance size, feature distribution, and tasks are captured in the experiment.    

\mypar{Future Work.}
%In the following, we discuss related applications of orbital labeling.
The most promising future work direction is the integration of interactivity, e.g., panning and zooming.\del{ Firstly, panning and zooming are common interactions for data.} Here, it would be interesting to investigate orbital labeling with regard to the stability of leader lines. While our heuristics can quickly recompute a crossing-free labeling after such an interaction, there is no guarantee that the order of labels remains the same. Hence, leader lines could suddenly connect to a different point on the boundary.

% Our orbital labeling can also produce thematic maps. For example, one can use it to display statistics. Given a set of point features (e.g., cities on a map) and a value for each of the point features (e.g., the number of citizens), one can scale the lengths of the labels proportional to the statistic value. Then, the labeling is similar to a necklace map or donut chart.

%Further, our orbital labeling enables us to display information of two categories by combining it with internal labeling. For example, for a map, the county labels that relate to polygons can be placed internally and the labels of the cities externally.

\begin{figure}[t]
    \centering
    \includegraphics[width=0.34 \linewidth]{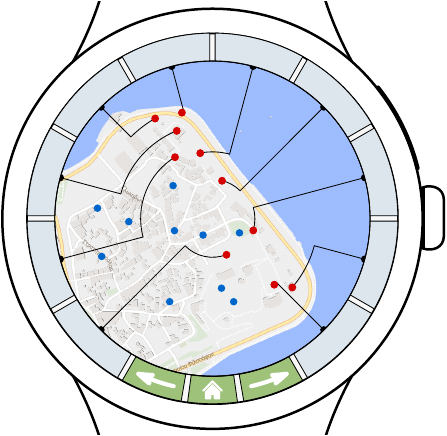}
    \caption{Pagination on the bottom is used to integrate the navigation and/or other system functions. The arrows change the set of labeled features, e.g., from the red to the blue point features.}
    \label{fig:enter-label}
\end{figure}

Secondly, the number of feature points to label can exceed the number of legible labels. Internal labeling has handled this by only showing a subset of (most important) labels or moving individual labels slightly.
We would like to highlight a strategy for handling a larger number of point features in external labeling. Gedicke et al.~\cite{GedickeBNH21} introduced three approaches for external labeling on rectangular small screens: the labels are \textbf{(1)} shown on partitions of features on \emph{multiple pages}, \textbf{(2)} ordered and the user can slide them along the label region, \textbf{(3)} aggregated into \emph{stacks}, where each label corresponds to multiple features and the user can flip through them.   
Combining our approach with theirs is a promising prospect. For example, combining it with the multi-page variant seems to be straightforward. Each labeling of a set can be computed independently of the others. We can place a button at the bottom of the labeling region, allowing the user to change the page; see \autoref{fig:enter-label}.

%Smartwatches often have their navigation buttons at the bottom border of the screen. We can easily exclude this part of the label region and combine our orbital labeling with the existing button design. This results in an aesthetically pleasing design for circular displays and the user is already accustomed to the interaction patterns.
 
\section{Conclusion} \label{sec:conclusion}

This work investigates the clarity and computational performance of orbital labeling, a recent external labeling technique designed for circular boundaries. We identify relevant orbital boundary labeling variants in a visualization context. Moreover, we present algorithms to compute several of the discussed variants of orbital labeling, which differ in label sizes and leader styles. We evaluate our algorithms quantitatively and demonstrate that our heuristics perform well in terms of runtime and solution quality. Furthermore, we conducted a user study with 54 participants, in which both leader styles performed equally well in terms of accuracy. However, straight leaders have significantly faster response times.

\acknowledgments{This research was partially funded by the Vienna Science and Technology Fund (WWTF) [10.47379/ICT19035].
S.T. was funded by the NWO Gravitation project NETWORKS under grant no. 024.002.003. 
J.W. was funded by the Dutch Research Council (NWO) under project number VI.Vidi.223.137.
}

\bibliographystyle{abbrv-doi}

\bibliography{bibliography}

\clearpage
\newpage
\appendix

\section{Runtime of the Computational Experiment}

\begin{figure}[t]
    \centering
    \includegraphics[width=\linewidth]{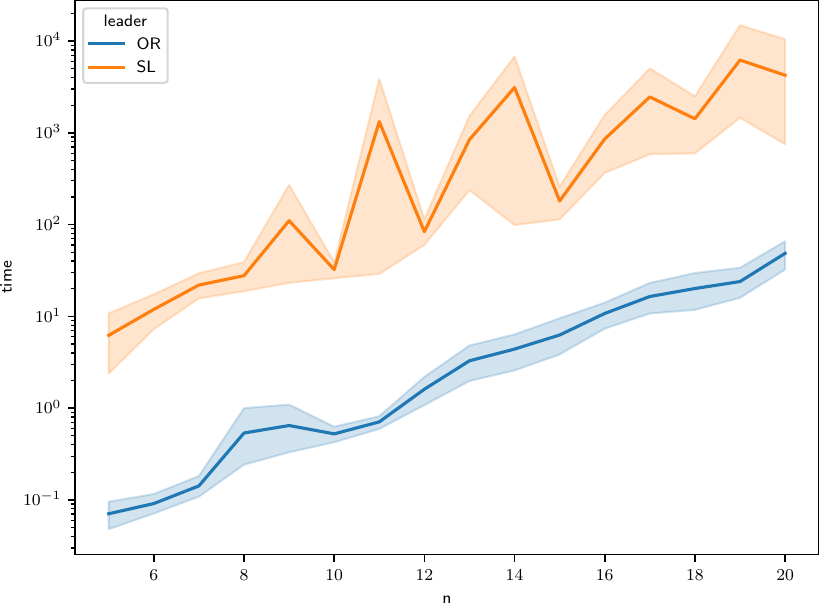}
    \caption{Runtime of the MIP and QIP models to compute OR- and SL-leader in seconds. SL-leader takes considerably longer to compute, and even the smallest instances already take around 10 seconds to compute. Additionally, both curves show exponential growth.}
    % $\TLL_r$ of SL- and OR-leaders.
    \label{fig:run_opt}
\end{figure}

\begin{figure}[t]
    \centering
    \includegraphics[width=\linewidth]{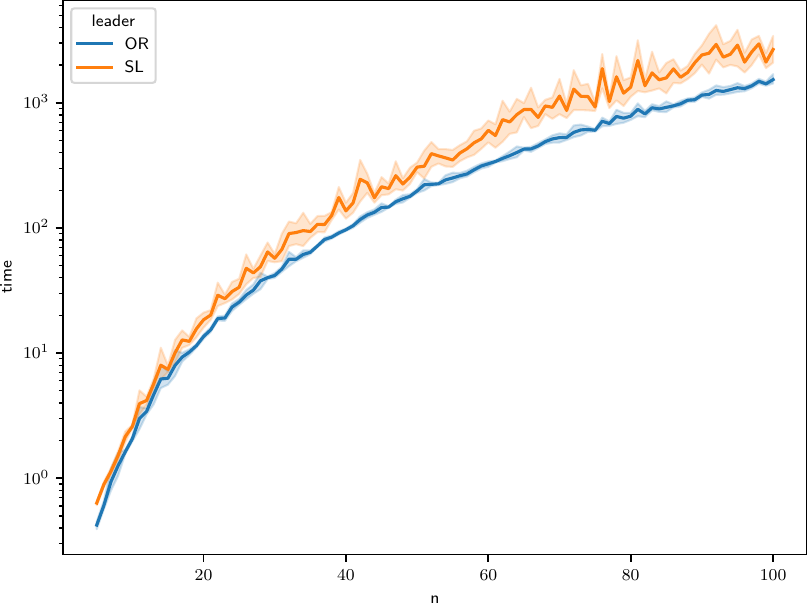}
    \caption{Runtime of the heuristic to compute OR- and SL-leader in milliseconds. Runtime behavior is similar for both leader types and exhibits linear growth.}
    % $\TLL_r$ of SL- and OR-leaders.
    \label{fig:run_heur}
\end{figure}

We present two additional figures of our computational experiment. In \Cref{fig:run_opt}, the runtime of the MIP and QIP models is presented, while \Cref{fig:run_heur} shows the runtime of the heuristics. \rev{The jagged runtime behavior of SL-leaders can be attributed to the use of a non-convex QIP formulation. In contrast to the linear MIP used for OR-leaders, small changes in instance size or geometry can significantly alter the branch-and-bound search tree of the QIP solver, leading to large runtime fluctuations even for similar values of $n$.}

\section{Optimal Labeling Algorithms}
\label{sec:algorithms_appendix}

We briefly introduce the orbital boundary labeling problem as stated in the work by Bonerath et al.~\cite{BonerathNTWW24} and discuss which variants have the most relevance in a visualization context.

\begin{figure}
	\centering
	\includegraphics[page=5,width=0.5\linewidth]{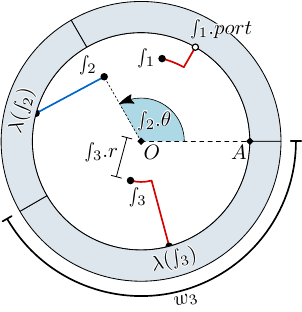}
	\caption{Illustration of the notation used in this paper. Note that both OR-leaders (red) and SL-leaders (blue) are illustrated, while any labeling we consider will exclusively contain one or the other.}
	\label{fig:notation_app}
\end{figure}

\mypar{Problem Description.}
The notation is illustrated in Figure~\ref{fig:notation_app}.
We are given a set $\mathcal{F} = \{\site_1, \dots, \site_n\}$ of point features.
The features are located within a disk with center~$O$, boundary~$B$ and radius~$R$.
Whenever we specify coordinates, we consider~$O$ to be the center of the coordinate system and we call the intersection point of a horizontal line starting at $O$ with $B$ the \emph{anchor}~$A$.
We assume that there is always one label whose circular arc along $B$ starts at $A$.
\revOld{Note that any other choice of $A$ would work, e.g., the bottom or top.}
A feature $\site_i$ is given in polar coordinates, i.e., the distance $\site_i.\distance$ from $\site_i$ to $O$ and the counter-clockwise angle $\site_i.\anglec = \measuredangle AO\site_i$.
%(see also Figure~\ref{fig:notation}). 
We will write $\vert B \vert$ for the length of~$B$.
A point feature has an associated label $\lambda(\site_i)$.
A label $\lambda(\site_i)$ is a circular arc along~$B$ of length~$w_i$, s.t., $\vert B\vert = \sum_{i=1}^n w_i$.
The output is a set of $n$ ports $\site_1.port, \ldots, \site_n.port$, s.t., the $\lambda(\site_i)$ can be placed with $\site_i.port$ at its center, all labels are pairwise interior-disjoint, and every $\site_i$ can be connected to the port at the center of $\lambda(\site_i)$ with a leader curve of a certain style.
In any valid solution, all leader curves have to be pairwise non-intersecting, and a solution is optimal if the total length of all leader curves is minimal over all valid solutions.
We refer to this as a \emph{leader length-optimal labeling}.
\revOld{Such an optimal solution has the advantage of requiring less ink, thus reducing the clutter.}

% properties of a labeling like 
Note that small gaps can be introduced between labels when rendering the labeling.
Finally, we assume a variant of general position for the placement of feature points, specifically for any pair of points $\site_i, \site_j$ we assume $\site_i.\distance \neq \site_j.\distance$. 
This assumption is made to avoid ambiguity between leaders and can be achieved by small perturbations to the feature points.
% This is a sensible assumption, as otherwise it is difficult to unambiguously associate leaders with two features at the same position.

\mypar{Variants.}
The four variants we consider are characterized by their leader style (SL vs.\ OR) and their label sizes (uniform vs.\ non-uniform).
%of the above problem definition, depending on two (binary) variables.
%First, we investigate two different leader styles: straight-line SL- and orbital-radial OR-leaders.
A \emph{straight-line} (SL) leader for a point feature~$\site_i$ is a line segment with endpoints~$\site_i$ and~$\site_i.port$, and its length is the Euclidean distance between these two endpoints.
In contrast, an \emph{orbital-radial} (OR) leader consists of an orbital circular arc concentric with $B$ starting at $\site_i$ and ending at the start point of a radial line segment (its supporting line contains $O$), which ends at~$\site_i.port$.
%Note that for a fixed position of~$\site_i$ and~$\site_i.port$ there are exactly two possible OR-leader.
The length of an OR-leader is the sum of the length of the circular arc segment and that of the radial line segment.
% Second, we distinguish variants based on the sizes~$w_i$ of the labels.
While SL-leaders for a chosen port are unique, OR leaders could run clockwise or counterclockwise; 
to avoid long leaders
% since we are minimizing leader length 
we always use the shorter of the two possibilities.
Label sizes are distinguished as follows.
If all labels have the same size $w_i = \vert B\vert / n$, then we say the labels are \emph{uniform}, otherwise they are \emph{non-uniform}.

We describe the full model of the mixed and quadratic integer program here. The same assumptions as in the main paper apply. 

% First, we introduce preliminaries for the problem on an abstract level. We assume that we are given a collection of sites $\sites = {\site_1, \dots, \site_n}$ where $n$ is the total number of sites. Let $O$, $B$ and $R$ be the center/origin of the coordinate system, the boundary of the circle and the radius of the circle respectively.\textcolor{red}{$R$ could simply be 1}
% We denote with $|B|$ the length of $B$.
% The Cartesian coordinates of a site are expressed as $\site_i.x$ and $\site_i.y$.
% Note that these coordinates can equivalently be expressed as polar coordinates consisting of the clockwise angle $\site.\angl$ between the point and the ray starting at $O$ and extending horizontally to the right  and a distance $\site.r$ from $O$.
% We frequently use the concept of polar coordinates in this section.
% Additionally, we assume that each site $\site_i$ has a weight $w_i$ assigned.
% Let $W = \sum_{i=1}^n w_i$.
% The label of site $\site_i$ is a radial arc of length $|B|\cdot w_i/W$ along $B$.
% We refer to its two endpoints as the clockwise and counter-clockwise endpoint (relative to $O$).
% We will refer to the port of the label of a site $\site_i$ as $\site_i.port$.
% Note that a port can also be expressed as a point in polar coordinates.
% All algorithms assume that anchor point is given, i.e., in any valid solution there is one label (the first label) whose clockwise endpoint has polar coordinates $(0, R)$.
% Note that the existence of this anchor point implies that a total order on the labels completely characterizes a solution.

We are able to formulate a mixed integer program, for computing optimal OR-labeling since crossings between OR-leaders can be determined combinatorially and the length of a OR-leader can be expressed as a linear function of the polar coordinates of its site and port.
However for straight line leaders we find ourselves in the need of computing the euclidean length of a line segment to determine an optimal solution and the mathematical programming model to compute an SL-labeling is a non-convex quadratic program.

\subsection{The OR-model.}
% \textcolor{red}{Short version:}
% We sketch the construction of the \textsf{MIP}.
% A full description can be found in Section~\ref{app:MIP}.
% The \textsf{MIP} uses a set of binary variables to encode a transitive order on all ports.
% Then the angular component of the polar coordinate of a port is represented with a real-valued variable.
% Since a labeling is completely characterized by the order on the labels, the port angle variables can be set based on the binary variables.
% Next the length of a leader in such a labeling can be calculated based on the port angle variable and the angle of a point feature (given in the input).
% The sum over all of these angle differences is minimized to obtain a leader length minimal labeling.
% Finally we can determine combinatorially if two leaders cross solely based on the angle variables of their ports; planarity can be encoded using binary indicator variables.

% We describe a mixed integer model, which computes leader length minimal OR labelings.
For a given instance with point features in a disk (given in polar coordinates), a solution is characterized by the position of the ports of each point feature $\site_i$.
The angle component of the polar coordinate of $\site_i.port$ is represented in the \textsf{MIP} as $\portangle_i$; note that we only require the angle, since the distance to $O$ is always $R$.
% With these variables we can formulate the objective function as follows.

% \[ \text{Minimize: } \sum \limits_{i=1}^{n} \left(\portangle_i\cdot \portdist_i\right) \]

We need to encode a total order on all labels.
Since this order is transitive, we use a standard encoding of transitivity in \textsf{MIP}s.
To do so we introduce a binary variable $a_{i,j}$, which encodes that (starting at the anchor point and proceeding counter-clockwise along $B$) $\lambda(\site_i)$ appears before label $\lambda(\site_j)$.
To guarantee transitivity, we subject these variables to the following constraint.

\[ \forall 1\leq i<j<k\leq n: 0 \leq a_{i,j} + a_{i,k} - a_{i,k}\leq 1\]

The set of all transitivity variables implies an order of all labels and therefore a labeling.
We now enforce that the $\portangle$ variables are set to the values corresponding to this solution by adding the following constraints.

\[ \forall 1\leq i \leq n: \portangle_i = w_i/2 + \sum \limits_{j=1}^{n} (a_{j,i}\cdot w_j)\]

We can now measure the angle that the circular arc of the leader of $\site_i$ spans based on $\portangle_i$.
This value is represented by a real valued variable $\delta_i$, which is constrained as follows.

\[ -\delta_i \leq \site_i.\anglec - \portangle_i + b_1 \cdot 2\pi - b_2 \cdot 2\pi \leq \delta_i \]

With the $\delta$ variables properly constrained, we can set up the objective function of our \textsf{MIP}.

\[ \text{Minimize: } \sum_{i=1}^{n} \delta_i\cdot\site_i.\distance\]

What remains is to enforce that no leaders cross.
To do so we can leverage the fact that a crossing between OR-leaders of two point features $\site_i$ and $\site_j$, where $\site_i.\distance < \site_j.\distance$, can be determined combinatorially solely based on $\portangle_i$, $\portangle_j$ and $\site_j.\anglec$.
We first set for every point feature $\site_i$ the value of an indicator variable $c_i$ determine which of the two values $\site_i.\anglec$ and $\portangle_i$ is larger.

\begin{align*}
\forall 1\leq i \leq n: 0 &\geq \portangle_i - \site_i.\anglec - c_i \cdot 2\pi\\
\forall 1\leq i \leq n: 0 &\leq \portangle_i - \site_i.\anglec - (1-c_i) \cdot 2\pi
\end{align*}

Second, depending on the value of $c_i$, we set a binary variable $d_i$, which indicates if the circular arc of a leader crosses the line segment $OA$ from the center to the anchor.

\begin{align*}
\forall 1\leq i \leq n: 0 &\leq \site_i.\anglec - \portangle_i - \pi + c_i \cdot 2\pi + d_i \cdot 2\pi\\
\forall 1\leq i \leq n: 0 &\geq \site_i.\anglec - \portangle_i - \pi - c_i \cdot 2\pi - (1-d_i) \cdot 2\pi\\
\forall 1\leq i \leq n: 0 &\leq \portangle_i - \site_i.\anglec - \pi + (1-c_i) \cdot 2\pi + d_i \cdot 2\pi\\
\forall 1\leq i \leq n: 0 &\geq \portangle_i - \site_i.\anglec - \pi - (1-c_i) \cdot 2\pi - (1-d_i) \cdot 2\pi
\end{align*}

With the two binary indicator variables $c_i$ and $d_i$ set, we now create a final set of binary variables $e^1_{i,j}$ and $e^2_{i,j}$.
We set this variable depending on the $c$ and $d$ variables as follows.

\begin{align*}
    \forall 1 \leq i < j \leq n: 0 &\geq \portangle_i - \site_j.\anglec - c_i\cdot2\pi - e^1_{i,j}\cdot2\pi\\
    \forall 1 \leq i < j \leq n: 0 &\leq \portangle_i - \site_j.\anglec + c_i\cdot2\pi + (1-e^1_{i,j})\cdot2\pi\\
    \forall 1 \leq i < j \leq n: 0 &\geq \portangle_j - \portangle_i - c_i\cdot2\pi - e^2_{i,j}\cdot2\pi\\
    \forall 1 \leq i < j \leq n: 0 &\leq \portangle_j - \portangle_i + c_i\cdot2\pi + (1-e^2_{i,j})\cdot2\pi\\
    \forall 1 \leq i < j \leq n: 0 &\geq \portangle_i - \portangle_j - (1-c_i)\cdot2\pi - e^1_{i,j}\cdot2\pi\\
    \forall 1 \leq i < j \leq n: 0 &\leq \portangle_i - \portangle_j + (1-c_i)\cdot2\pi + (1-e^1_{i,j})\cdot2\pi\\
    \forall 1 \leq i < j \leq n: 0 &\geq \site_j.\anglec - \portangle_i - (1-c_i)\cdot2\pi - e^2_{i,j}\cdot2\pi\\
    \forall 1 \leq i < j \leq n: 0 &\leq \site_j.\anglec - \portangle_i + (1-c_i)\cdot2\pi + (1-e^2_{i,j})\cdot2\pi
\end{align*}

With this we are finally prepared to enforce non-crossing leaders for any pair of point features.
This is done through the following constraints.

\begin{align*}
    \forall 1 \leq i < j &\leq n: 1 \leq e^1_{i,j} + e^1_{i,j} + (1-d_j)\\
    \forall 1 \leq i < j &\leq n: 1 \geq e^1_{i,j} + e^1_{i,j} - (1-d_j)\\
    \forall 1 \leq i < j &\leq n: 0 \leq e^1_{i,j} + e^1_{i,j} + d_j\\
    \forall 1 \leq i < j &\leq n: 0 \geq e^1_{i,j} + e^1_{i,j} - d_j
\end{align*}

\subsection{The SL-model. }
% \textcolor{red}{Short version:}
% The \textsf{QIP} shares some DNA with the \textsf{MIP}.
% We again only sketch this \textsf{QIP} here and refer to Section~\ref{app:QIP} for a full description.
% In particular, the port positions are represented and transitivity is enforced the same way.
% However length of a leader now needs to be determined as the euclidean distance between a feature point and its port.
% by adding the following (non-linear) constrains we can optimize the total leader length in the objective function.

% \[ \forall 1\leq i \leq n: \gamma_i \geq \sqrt{(\site_i.\distance)^2 + R^2 - 2\cdot\site_i.\distance\cdot R \cos(\site_i.\anglec - \portangle_i)} \]
% \[ \text{Minimize: } \sum_{i=1}^{n} \gamma_i\]

% It remains to enforce planarity.
% This is done via careful case distinction.
% For two leaders $p_1q_1$ and $p_2q_2$ consider the supporting lines $\ell_1$ and $\ell_2$ respectively.
% We check if (i) $p_1$ and (ii) $q_1$ are on different sides of $\ell_2$ and if (iii) $p_2$ and (iv) $q_2$ are on different sides of $\ell_1$.
% If and only if all four checks are true, the leaders cross.
% This is again encoded using non-linear constrains and binary indicator variables.

When computing a SL-label we still want to minimize leader length, however the length of a leader in this style corresponds to the euclidean distance between a point feature and its port, which is not a linear function.
We therefore formulate a \textsf{QIP}.
The ports are again represented with a set of $\portangle$ variables, which are ensured to be correctly placed with transitivity constrains.
However their length is measured differently, specifically we use the following constrains.

\[ \forall 1\leq i \leq n: \gamma_i \geq \sqrt{(\site_i.\distance)^2 + R^2 - 2\cdot\site_i.\distance\cdot R \cos(\site_i.\anglec - \portangle_i)} \]

It is worth noting that neither the trigonometric function nor the square root are computed exactly by the software used to implement this model, but are instead approximated with piecewise linear functions; we refer to the Gurobi documentation\footnote{\href{https://www.gurobi.com/documentation/current/refman/java_model_agc_pow.html}{Power constraints}, \href{https://www.gurobi.com/documentation/current/refman/py_model_agc_sin.html}{sine constraints} and \href{https://www.gurobi.com/documentation/current/refman/java_model_agc_cos.html}{cosine constraints} specifically.} for more detail.
The $\delta$ values can then be used straight-forwardly in the objective function.

\[ \text{Minimize: } \sum_{i=1}^{n} \gamma_i\]

It is left to describe how crossings are prevented.
We do this by modeling four checks.
To ease description we use Cartesian coordinates, which can be obtained based on the polar coordinates used to represent the point features and ports.
For two leaders $p_1q_1$ and $p_2q_2$ consider the supporting lines $\ell_1$ and $\ell_2$ respectively.
We check if (i) $p_1$ and (ii) $q_1$ are on different sides of $\ell_2$ and if (iii) $p_2$ and (iv) $q_2$ are on different sides of $\ell_1$.
If and only if all four checks are true, the leaders cross.
The checks are formalized in the following set of constraints setting the four binary variables $f^i$ for $i\in\{1, 2, 3, 4\}$.

\begin{align*}
    (q_2.y - p_1.y) &(p_2.x - p_1.x) \geq\\
    &(p_2.y - p_1.y) (q_2.x - p_1.x) - f^1_{i,j}\cdot 2\pi\\
    (q_2.y - p_1.y) &(p_2.x - p_1.x) \leq\\ 
    &(p_2.y - p_1.y) (q_2.x - p_1.x) + (1-f^1_{i,j})\cdot 2\pi\\
    (q_2.y - q_1.y) &(p_2.x - q_1.x) \geq\\
    &(p_2.y - q_1.y) (q_2.x - q_1.x) - f^2_{i,j}\cdot 2\pi\\
    (q_2.y - q_1.y) &(p_2.x - q_1.x) \geq\\ 
    &(p_2.y - q_1.y) (q_2.x - q_1.x) + (1-f^2_{i,j})\cdot 2\pi\\
    (p_2.y - p_1.y) &(q_1.x - p_1.x) \geq\\
    &(p_1.y - p_1.y) (q_2.x - p_1.x) - f^3_{i,j}\cdot 2\pi\\
    (p_2.y - p_1.y) &(q_1.x - p_1.x) \geq\\ 
    &(p_1.y - p_1.y) (q_2.x - p_1.x) + (1-f^3_{i,j})\cdot 2\pi\\
    (q_2.y - p_1.y) &(q_1.x - p_1.x) \geq\\
    &(q_1.y - p_1.y) (q_2.x - p_1.x) - f^4_{i,j}\cdot 2\pi\\
    (q_2.y - p_1.y) &(q_1.x - p_1.x) \geq\\ 
    &(q_1.y - p_1.y) (q_2.x - p_1.x) + (1-f^4_{i,j})\cdot 2\pi
\end{align*}

We set another set of two binary variables $g^i$ for $i\in\{1, 2\}$.

\begin{align*}
    g^1_{i,j} &\leq f^1_{i,j} + f^2_{i,j}\\
    g^1_{i,j} &\geq f^1_{i,j} - f^2_{i,j}\\
    g^1_{i,j} &\geq f^2_{i,j} - f^1_{i,j} \\
    g^1_{i,j} &\leq 2 - f^1_{i,j} - f^2_{i,j}\\
    g^2_{i,j} &\leq f^3_{i,j} + f^4_{i,j}\\
    g^2_{i,j} &\geq f^3_{i,j} - f^4_{i,j}\\
    g^2_{i,j} &\geq f^4_{i,j} - f^3_{i,j} \\
    g^2_{i,j} &\leq 2 - f^3_{i,j} - f^4_{i,j}
\end{align*}

And finally, the two binary variables are used in the final constraint.

\[ g^1_{i,j} + g^2_{i,j} \leq 1\]

\section{Additional Orbital Labeling Examples and Analysis}

\begin{figure}[t]
    \centering
    \includegraphics[width=\linewidth]{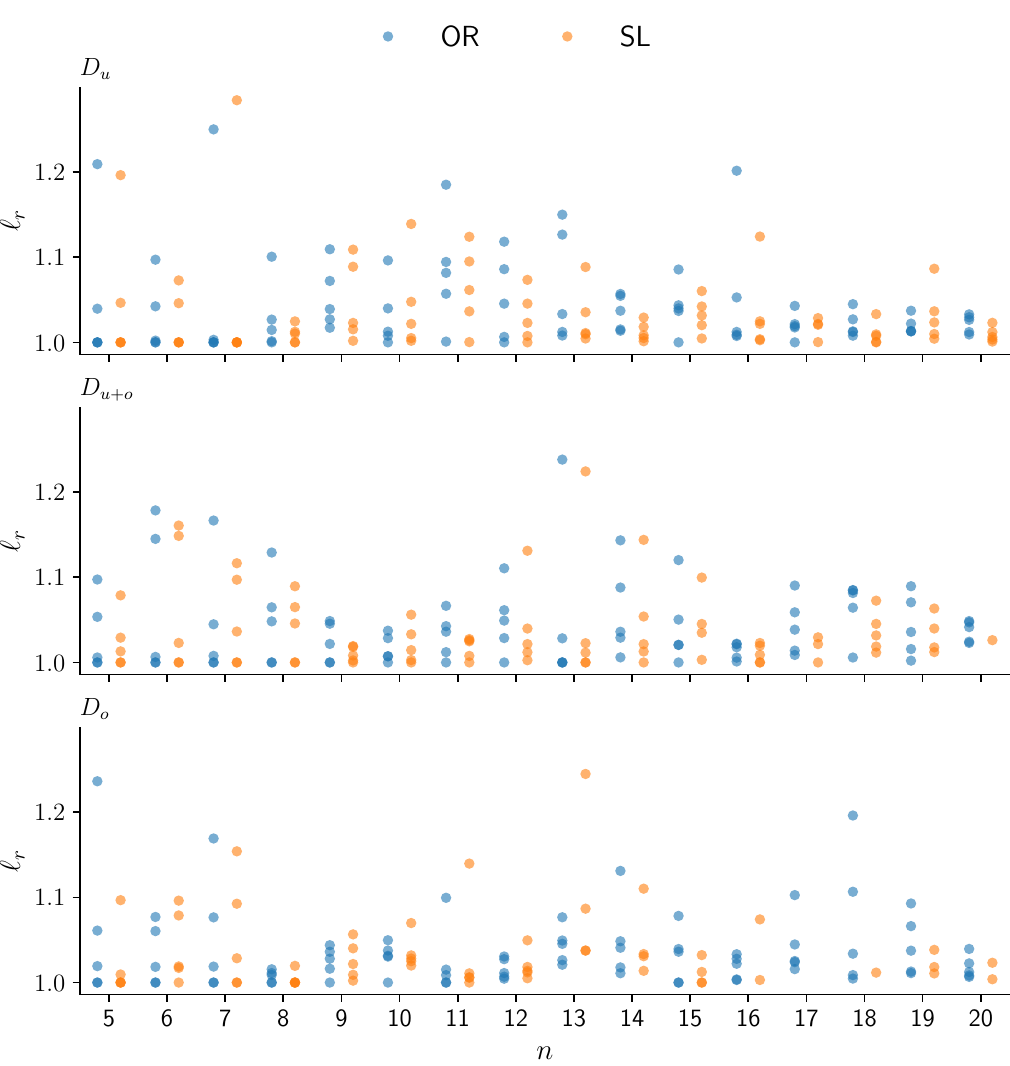}
    \caption{Solution quality of the heuristic split by instance classes uniform ($D_u$), uniform and off-center $D_{u+o}$, and off-center $D_o$. The plots indicate no obvious differences between classes.}
    % $\TLL_r$ of SL- and OR-leaders.
    \label{fig:qual_class}
\end{figure}

In this section, we present additional examples of orbital labeling to show the best and worst cases of the heuristic, as well as additional real-world instances. Additionally, we provide an extended analysis of the solution quality of the heuristic. We split the solutions by instances and perform the same analysis. \Cref{fig:qual_class} shows the results. The plot shows no significant differences between the classes.

\begin{figure}[t]
    \centering
    \begin{subfigure}{0.8\linewidth}
        \centering
        \includegraphics[width=\linewidth]{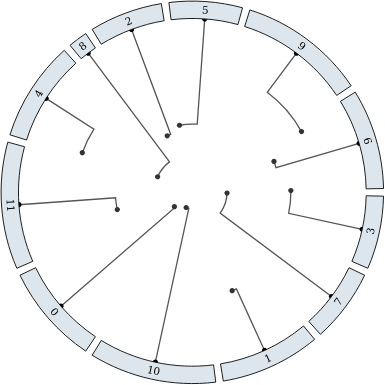}
        \subcaption{}
    \end{subfigure}
    \hfill
    \begin{subfigure}{0.8\linewidth}
        \centering
        \includegraphics[width=\linewidth]{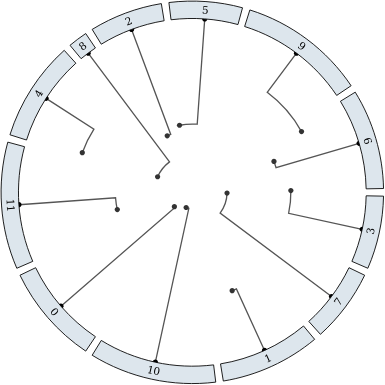}
        \subcaption{}
    \end{subfigure}
    \caption{One example where the heuristic found an optimal solution for OR-leaders with 12 features. The illustrated instance is ``uniform\_12\_3.json''.}
    \label{fig:opt_sol}
\end{figure}

We first investigate an example of the heuristics finding an optimal solution. Overall, in 19.5\% of all instances, either heuristic returned an optimal solution. An example of OR-leaders is given in \Cref{fig:opt_sol}. 

\begin{figure}[t]
    \centering
    \begin{subfigure}{0.8\linewidth}
        \centering
        \includegraphics[width=\linewidth]{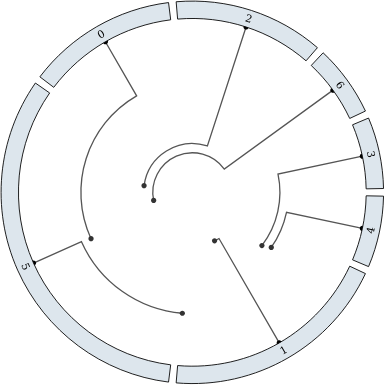}
        \subcaption{}
    \end{subfigure}
    \hfill
    \begin{subfigure}{0.8\linewidth}
        \centering
        \includegraphics[width=\linewidth]{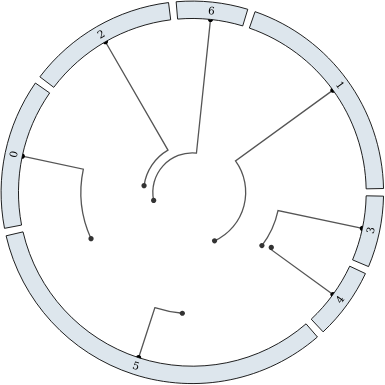}
        \subcaption{}
    \end{subfigure}
    \caption{The instance in our dataset where the heuristic found the overall worst solution for OR-leaders with 7 features. In (a) shows the heuristic solution and (b) the optimal solution by the MIP. The total leader length of the heuristic solution is 1.28 times the length of the optimal solution The illustrated instance is ``uniform\_7\_0.json''.}
    \label{fig:worst}
\end{figure}

Regarding the worst solution, one instance in the dataset resulted in a heuristic solution with 1.28 times the total leader length of the optimal solution. The instance is shown in \Cref{fig:worst}. Even though there are only 7 features, the poor performance is driven by the choice of how features' 0' and '5' are labeled. The reason for this is that the heuristic labels '1' with a short leader line, thus requiring '0' and '5' to have long leader lines.

\begin{figure*}[t]
    \centering
    \begin{subfigure}{.45\linewidth}
        \centering
        \includegraphics[width=\linewidth]{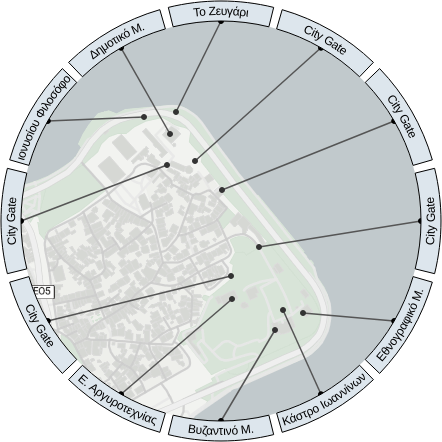}
        \subcaption{}
    \end{subfigure}
    \quad
    \begin{subfigure}{.45\linewidth}
        \centering
        \includegraphics[width=\linewidth]{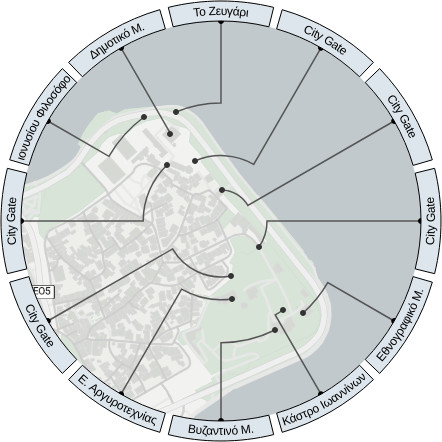}
        \subcaption{}
    \end{subfigure}
    \quad
    \begin{subfigure}{.45\linewidth}
        \centering
        \includegraphics[width=\linewidth]{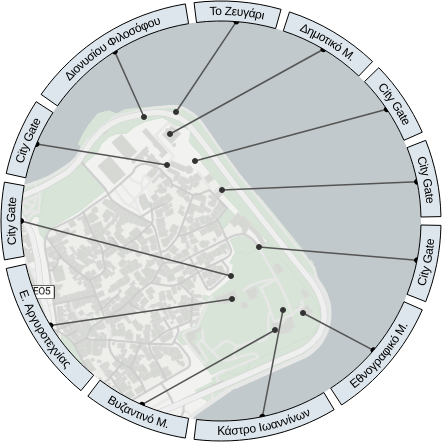}
        \subcaption{}
    \end{subfigure}
    \quad
    \begin{subfigure}{.45\linewidth}
        \centering
        \includegraphics[width=\linewidth]{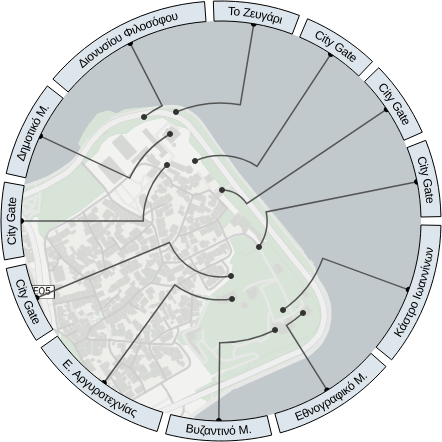}
        \subcaption{}
    \end{subfigure}
    \caption{Historic district of a city in Greece. All 12 features are historically important buildings. The top row shows solutions with uniform labeling, while the bottom row shows heuristic solutions with non-uniform labeling. Uniform labeling results in overall shorter total leader length, however, it is unlikely to fit another label without decreasing font size.}
    \label{fig:ex_2}
\end{figure*}

We show another real-world example in \Cref{fig:ex_2}. This shows the historic district of a city in Greece. All 12 features are historically important buildings. The top row shows solutions with uniform labeling, while the bottom row shows heuristic solutions with non-uniform labeling. Due to the uniform distribution of features, the total leader length for uniform labels is less than that for non-uniform labels. However, some of the longer labels barely fit the available space and adding more labels is likely impossible without decreasing the font size.

\section{SL-leaders for Uniform Labels can have no Planar Solution} \label{sec:counterexample}

\begin{figure*}[t]
    \centering
    \begin{subfigure}{.3\linewidth}
        \centering
        \includegraphics[width=\linewidth,page=1]{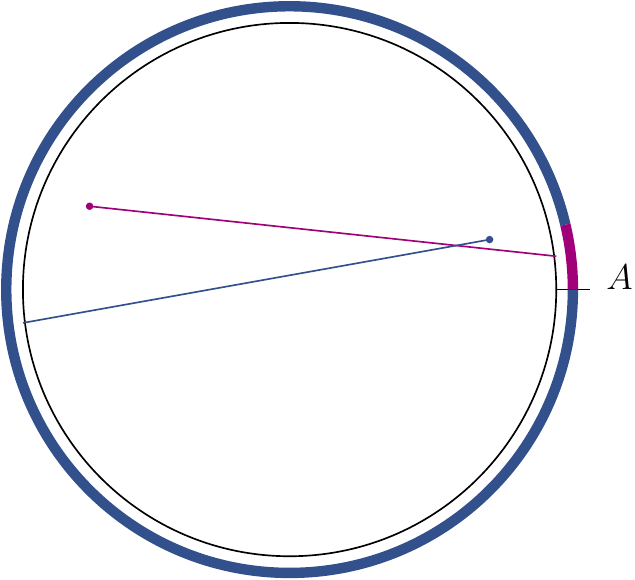}
        \subcaption{}
        \label{fig:counter_example_sketch1}
    \end{subfigure}
    \quad
    \begin{subfigure}{.3\linewidth}
        \centering
        \includegraphics[width=\linewidth,page=2]{figures/counter.pdf}
        \subcaption{}
    \end{subfigure}
    \\
    \begin{subfigure}{.3\linewidth}
        \centering
        \includegraphics[width=\linewidth,page=3]{figures/counter.pdf}
        \subcaption{}
        \label{fig:counter_example_sketch2}
    \end{subfigure}
    \quad
    \begin{subfigure}{.3\linewidth}
        \centering
        \includegraphics[width=\linewidth,page=4]{figures/counter.pdf}
        \subcaption{}
    \end{subfigure}
    \quad
    \begin{subfigure}{.3\linewidth}
        \centering
        \includegraphics[width=\linewidth,page=5]{figures/counter.pdf}
        \subcaption{}
    \end{subfigure}
    \quad
    \begin{subfigure}{.3\linewidth}
        \centering
        \includegraphics[width=\linewidth,page=8]{figures/counter.pdf}
        \subcaption{}
    \end{subfigure}
    \quad
    \begin{subfigure}{.3\linewidth}
        \centering
        \includegraphics[width=\linewidth,page=7]{figures/counter.pdf}
        \subcaption{}
    \end{subfigure}
    \quad
    \begin{subfigure}{.3\linewidth}
        \centering
        \includegraphics[width=\linewidth,page=6]{figures/counter.pdf}
        \subcaption{}
    \end{subfigure}
    \caption{ (a)-(b) illustrates a counter example with two features. (c)-(h) illustrates all configurations for an example with three features and restricted label size.}
    \label{fig:counter_example_sketch}
\end{figure*}

Unfortunately, the use of an anchored boundary introduces restrictions on the solution space for SL-leaders with non-uniform labels. As a result, there exist instances where no planar solution is possible. \Cref{fig:counter_example_sketch} illustrates two such counterexamples. In the first case (\Cref{fig:counter_example_sketch1}), even just two features can cause crossings if one label occupies an excessively large portion of the boundary. Moreover, even if we impose reasonable constraints on label size, such as limiting each label to at most half of the boundary's circumference, planar solutions may still be unattainable under certain feature configurations, as shown in \Cref{fig:counter_example_sketch2}. We conjecture that even if the label size is further constrained, counterexamples can be constructed.

\end{document}